\documentclass[a4paper,fleqn]{cas-dc}

\usepackage[utf8]{inputenc}
\usepackage{lmodern}
\usepackage{pifont}
\providecommand{\tightlist}{\setlength{\itemsep}{0pt}\setlength{\parskip}{0pt}}

\def\tsc#1{\csdef{#1}{\textsc{\lowercase{#1}}\xspace}}
\tsc{IDS}
\tsc{CII}

\ExplSyntaxOn
\cs_gset:Npn \__first_footerline:
  {
    \group_begin:
    \small
    \sffamily
    \__short_authors: : ~ Preprint
    \group_end:
  }
\ExplSyntaxOff

\begin{document}
\let\WriteBookmarks\relax
\def\floatpagepagefraction{1}
\def\textpagefraction{.001}

\shorttitle{Adaptive IDS with Transformers, Continual Learning, and Adversarial Investigation}

\shortauthors{Ariffin et~al.}

\title[mode=title]{Adaptive Intrusion Detection System using Transformer-Based Neural Networks and Continual Learning Approach with Adversarial Investigation}

\author[1,3]{Azizi Ariffin}[orcid=0000-0002-8183-0457]
\credit{Conceptualization, Grant Acquisition, Methodology, Software, Writing -- original draft}

\author[1,4]{Afif Haris}
\credit{Investigation, Data curation, Writing -- original draft}

\author[1]{Faiz Zaki}[orcid=0000-0001-7030-4593]
\credit{Investigation, Validation}

\author[1,2]{Hazim Hanif}
\credit{Investigation, Validation}

\author[1]{Nor Badrul Anuar}[orcid=0000-0003-4380-5303]
\cormark[1]
\ead{badrul@um.edu.my}
\credit{Supervision, Grant Acquisition, Writing -- review \& editing}

\affiliation[1]{organization={Centre of Research for Cyber Security and Network (CSNET), Faculty of Computer Science and Information Technology, Universiti Malaya},
            city={Kuala Lumpur},
            country={Malaysia}}

\affiliation[2]{organization={Department of Software Engineering, Faculty of Computer Science and Information Technology, Universiti Malaya},
            city={Kuala Lumpur},
            country={Malaysia}}

\affiliation[3]{organization={Faculty of Computer and Mathematical Sciences, Universiti Teknologi MARA},
            city={Shah Alam},
            state={Selangor},
            country={Malaysia}}

\affiliation[4]{organization={Network Security Research Department, Malaysian Communications and Multimedia Commission},
            city={Cyberjaya},
            country={Malaysia}}

\cortext[1]{Corresponding author.}

\begin{abstract}
Network intrusion detection systems (IDS) trained on fixed traffic snapshots decay silently after deployment as threat distributions shift. Fine-tuning models on new attacks triggers catastrophic forgetting, while retraining from scratch is computationally infeasible. Replay-based continual learning counters this, but existing methods unrealistically confine benign traffic to a single early task and ignore the replay buffer as a potential attack surface. To address this, we present an adaptive IDS framework coupling a tabular transformer encoder with a class-balanced experience replay buffer that replays benign traffic at every update to stabilize decision boundaries. We introduce the class-instance incremental (CII) scenario where benign flows reappear alongside new attacks as a more faithful stress test, and probe the buffer with overt label-flipping and stealthy backdoor poisoning attacks. On the CICIDS2017 benchmark, our framework achieved 0.9994 accuracy under the traditional class-incremental setup and 0.9989 under CII, with negligible forgetting, drastically outperforming sequential fine-tuning (0.0052), EWC (0.0324), LwF (0.0699), and iCaRL (0.8770) baselines. While injecting benign traffic into every experience proves essential for preventing forgetting, the replay buffer introduces critical vulnerabilities. Label-flipping collapses the model entirely (0.0053 accuracy at a 1\% budget), and the backdoor maintains >0.97 overall accuracy while driving the attack success rate on trigger flows to 95–100\%, evading standard monitoring. Ultimately, while a modest replay budget recovers near-joint-training performance, ensuring buffer integrity emerges as a strict operational requirement.
\end{abstract}


\maketitle

\section{Introduction}\label{introduction}

The attack surface that network defenders must police is expanding
faster than the models built to watch it. The CVE programme catalogued
48,185 new vulnerability disclosures in 2025, the largest annual total
on record (The MITRE Corporation, 2026), and each disclosure seeds fresh
exploit tooling, scanning campaigns, and malicious flow patterns that
arrive at the network edge within days. Network intrusion detection
systems (IDS) are the first line of defence for enterprise and
critical-infrastructure networks, and deep neural networks now
outperform classical signature-matching and hand-engineered statistical
methods on public benchmarks (Vinayakumar et al., 2019; Yin et al.,
2017; Shone et al., 2018). That advantage, however, is perishable. A
detector trained on a fixed snapshot of traffic decays silently after
deployment: shifts in traffic patterns degrade detection accuracy over
time (Shyaa et al., 2024), consequently, a model trained on one period's
traffic distribution will encounter attack classes from subsequent
periods that lie entirely outside its original training distribution (A
H et al., 2025; De Lange et al., 2022; Parisi et al., 2019). The cost of
that decay is measured in missed intrusions rather than in accuracy
points, which makes keeping a deployed detector current an urgent
operational problem rather than an academic one.

Retraining from scratch on the cumulative history, the obvious remedy,
is operationally infeasible for a security operations centre: the
accumulated history grows without bound, labelling is costly and often
adversarial, and each retraining window leaves attacks undetected for
its duration (Kim et al., 2022). The alternative, fine-tuning the
deployed model on new attack data, triggers catastrophic forgetting, the
phenomenon first reported by McCloskey and Cohen in connectionist
networks (McCloskey \& Cohen, 1989) and subsequently confirmed in modern
gradient-based neural networks (Goodfellow et al., 2013), in which new
learning overwrites previously learned decision boundaries. Three
threads of recent work apply continual learning to stabilise IDS models
against this failure. Regularisation-based methods such as EL-GNN
(Nguyen \& Park, 2025) penalise drift away from parameters that mattered
for earlier tasks, inheriting the formulation of Elastic Weight
Consolidation (Kirkpatrick et al., 2017); gradient-projection methods
such as SPIDER (Amalapuram et al., 2024) constrain updates to a subspace
orthogonal to earlier tasks; and memory-replay methods (A H et al.,
2025; Amalapuram et al., 2023) interleave a small buffer of past
examples during training, delivering the strongest empirical resistance
to forgetting under the severe class imbalance characteristic of
intrusion traffic. Two blind spots recur across all three threads.
First, they are evaluated almost exclusively under a class-incremental
scenario in which benign traffic appears in a single early experience,
although a production IDS sees benign as the dominant class in every
experience, every day; evaluations that hide this fact overstate
robustness to forgetting. Second, the replay buffer on which the
strongest methods depend has never been examined as an attack surface,
even though it persists between updates and is consulted at every
subsequent training step.

This paper presents an adaptive IDS framework that answers both blind
spots and the underlying architectural question of what the buffer
should preserve. The backbone is a tabular transformer encoder (Huang et
al., 2020), which applies multi-head self-attention over categorical
feature tokens so that the representation of one flow field, for example
a destination port, depends on the values of the others; this contextual
embedding matches the semantics of network-flow records better than the
static embeddings of a multi-layer perceptron. The encoder is coupled
with a class-balanced experience replay buffer that anchors the benign
class in every replay update, keeping the benign-attack decision
boundary well-posed at every update. Addressing the first blind spot, we
introduce the class-instance incremental (CII) scenario, in which benign
flows reappear in every experience alongside the new attack classes, and
argue that it should become the default evaluation for
continual-learning IDS research. Addressing the second, we mount two
poisoning attacks on the buffer under an explicit threat model, an overt
label-flipping attack and a stealthy backdoor that hides a
timing-feature trigger, and quantify the damage of each. The objective
is to establish whether a replay-based continual-learning IDS can absorb
new attack families without forgetting those it has already learned
under a production-faithful exposure pattern, and how far that stability
survives when the buffer itself is corrupted.

All claims are evaluated on the CICIDS2017 benchmark, partitioned into
four sequential experiences under both the standard class-incremental
(CI) scenario and the proposed CII scenario, with forgetting and
intransigence reported alongside accuracy and macro-F1. The framework
attains 0.9994 overall accuracy under CI and 0.9989 average per-task
accuracy under CII, against a sequential fine-tuning baseline that
collapses to 0.0052 accuracy, an EWC baseline at 0.0324, a
Learning-without-Forgetting baseline at 0.0699, and an iCaRL baseline
that partially recovers to 0.8770. Under attack, the picture splits by
objective: flipping the stored exemplar labels collapses the model
outright (accuracy from 0.9995 to 0.0053), whereas a backdoor that hides
a timing-feature trigger in the buffer holds overall accuracy above 0.97
while driving the attack success rate on trigger-carrying flows to 95 to
100\%, a silent corruption that an accuracy dashboard would miss.

Our five specific contributions are as follows:

\begin{itemize}
\tightlist
\item
  We demonstrate, on a public benchmark, that a sequentially fine-tuned
  tabular transformer IDS collapses to near-zero accuracy on earlier
  tasks after only four experiences.
\item
  We propose an adaptive IDS framework coupling a tabular transformer
  encoder with benign-anchored class-balanced experience replay,
  recovering near-joint-training performance with a modest memory
  budget.
\item
  We introduce the class-instance incremental (CII) evaluation scenario,
  show why the permissive CI scenario overstates robustness, and
  demonstrate that injecting a small amount of benign traffic into every
  experience during fine-tuning reduces catastrophic forgetting.
\item
  We provide a comprehensive evaluation of six continual-learning
  baselines (Na"ive sequential fine-tuning, Elastic Weight
  Consolidation, Learning without Forgetting, iCaRL, ER-Stratified, and
  ER-Balanced) on CICIDS2017 under both scenarios, reporting accuracy,
  macro-F1, forgetting, and intransigence in full.
\item
  We present an adversarial investigation of the replay buffer in a
  continual-learning IDS, quantifying two poisoning attacks, an overt
  label-flipping attack and a stealthy inter-arrival-time backdoor, on
  the stored exemplars at buffer budgets of 1 to 10\% under a common
  threat model.
\end{itemize}

The remainder of this paper is organised as follows. Section 2 surveys
deep-learning IDS, tabular transformers, continual learning, and their
intersection, and crystallises the research gap. Section 3 formalises
the problem, presents the framework, and defines the threat model.
Section 4 describes the experimental setup, Section 5 reports results
for both scenarios and the adversarial investigation, Section 6
discusses deployment implications, Section 7 outlines limitations and
future work, and Section 8 concludes.

\section{Related Work}\label{related-work}

\subsection{Deep learning for intrusion
detection}\label{deep-learning-for-intrusion-detection}

Deep learning entered intrusion detection through architectures
transplanted from vision and language. Convolutional and recurrent
networks applied to flow records (Yin et al., 2017) improved detection
rates over shallow classifiers on NSL-KDD and CICIDS benchmarks by
capturing local temporal patterns and higher-order feature interactions.
Autoencoder-based anomaly detectors learn a compact model of benign
traffic and flag outliers (Shone et al., 2018), an approach that avoids
the need for labelled attacks but tends to collapse under concept drift.
A common thread across these designs is that they treat a flow record
either as a fixed vector passed through dense layers or as a
pseudo-image reshaped for a CNN. Both treatments discard the fact that a
network flow is a structured, heterogeneous tuple whose fields carry
domain-specific meaning, and both ignore the situation in which the
label set changes over time. Few-shot prototype detectors such as the
dynamic prototype network of Chai et al.~(2023) adapt to new malware
classes from limited samples, but they assume a fixed feature backbone
and are not designed to accumulate classes across a long deployment.

\subsection{Transformer-Based
Architectures}\label{transformer-based-architectures}

The transformer architecture, introduced by Vaswani et al.~(2017),
replaced recurrence and convolution with multi-head self-attention,
allowing every token in a sequence to attend directly to every other
token and removing the sequential bottleneck of RNNs. This design
enabled parallel training at scale and has since become the dominant
backbone across natural language processing, computer vision, and
speech, where pre-trained variants such as BERT and GPT achieve
state-of-the-art results on a wide range of downstream tasks. The same
attention mechanism has more recently been adapted to structured data,
motivating the family of tabular transformers discussed below.

Flow records are tabular, and gradient-boosted decision trees have long
dominated tabular learning because they partition heterogeneous feature
spaces effectively and require little preprocessing. Their lack of
end-to-end differentiability and their inability to continue learning
from a stream make them awkward partners for continual IDS. The
transformer-based tabular models that have emerged in the last five
years close this gap. TabTransformer (Huang et al., 2020) applies
multi-head self-attention over learned embeddings of categorical
features, producing representations whose value depends on the row
context rather than on the feature index alone. FT-Transformer and SAINT
extend the idea by embedding numerical features as well and by
introducing inter-sample attention (Gorishniy et al., 2021; Somepalli et
al., 2021). TabNet (Arik \& Pfister, 2021) uses a sequential attention
mechanism with sparse feature selection. These models close the accuracy
gap with boosted trees on standard tabular benchmarks while retaining
gradient-based optimisation, which is exactly what a continual-learning
IDS needs.

\subsection{Continual learning and catastrophic
forgetting}\label{continual-learning-and-catastrophic-forgetting}

Catastrophic forgetting, first reported by McCloskey and Cohen (1989),
refers to the collapse in performance on previously learned tasks when a
neural network is updated on new data. Remedies fall into three
families. Regularisation-based approaches, including Elastic Weight
Consolidation (EWC) (Kirkpatrick et al., 2017), Synaptic Intelligence
(SI) (Zenke et al., 2017), and Learning without Forgetting (LwF) (Li \&
Hoiem, 2018), add a loss term that penalises parameter drift away from
values important for earlier tasks or that enforces output consistency
with an earlier snapshot. Replay-based approaches, including iCaRL
(Rebuffi et al., 2017), Gradient Episodic Memory and its averaged
variant (Lopez-Paz \& Ranzato, 2017; Chaudhry et al., 2019a), Experience
Replay with tiny episodic memories (Chaudhry et al., 2019b) and
class-balanced sampling (Chrysakis \& Moens, 2020), and Maximally
Interfered Retrieval (Aljundi et al., 2019), store a small subset of
prior examples and interleave them during new-task training. Dark
Experience Replay (Buzzega et al., 2020) unifies replay with soft-label
distillation, and meta-learning approaches such as MER balance transfer
and interference across tasks (Riemer et al., 2019). Architectural
approaches allocate new parameters for new tasks; they are difficult to
apply in a setting with strict inference-time budgets. The core tension
across all families is the stability--plasticity dilemma: preserve too
much and the model is too rigid to learn new attack patterns (high
intransigence); preserve too little and the model forgets.

\subsection{Continual learning applied to
IDS}\label{continual-learning-applied-to-ids}

Four recent works define the current state of continual-learning IDS.
Amalapuram et al.~(2023) propose ECBRS, an extension of class-balanced
reservoir sampling to severe imbalance, and PAPA, a Gaussian-mixture
approximation that accelerates maximally interfered sample retrieval;
they evaluate on KDDCUP'99 (Tavallaee et al., 2009), NSL-KDD,
CICIDS-2017/2018, UNSW-NB15, CTU-13, and the long-horizon AnoShift
benchmark, demonstrating that replay with buffer-management awareness is
essential under imbalance. SPIDER (Amalapuram et al., 2024) replaces
replay with gradient projection memory and achieves comparable detection
with only 20\% of labels, emphasising privacy and training-time
efficiency. TAMR (A H et al., 2025), the Elsevier Computer Networks
anchor of this work, introduces task-aware replay that scores each
buffered sample by its relevance to the current task rather than
replaying uniformly; it reports gains on five NIDS datasets while
lowering replay computational overhead. EL-GNN (Nguyen \& Park, 2025)
regularises a graph convolutional IDS by penalising drift of weights
identified as important for prior tasks, improving accuracy and F1 on
CIC-IDS2017 and UNSW-NB15 in a task-incremental setting. Our
contribution is complementary to all four. We do not propose a new
replay-selection rule; we retain a straightforward class-balanced buffer
and anchor the benign class in every replay update. Our architectural
backbone is a tabular transformer rather than a graph convolutional
encoder or a non-transformer MLP. Above all, we argue that none of the
four evaluates under a scenario that mirrors production exposure: in CI,
benign traffic is seen only in the first experience, which obscures the
most realistic interference pattern. Our class-instance incremental
scenario addresses exactly this gap.

Beyond these four anchors, the broader literature offers complementary
perspectives that motivate our design. Amalapuram et al.~(2022)
presented one of the earliest applications of continual learning to
anomaly-based NIDS, establishing that naive sequential fine-tuning
collapses under realistic traffic streams and that replay is the most
reliable counter-measure on imbalanced security data. Smith et
al.~(2024) propose adaptive memory replay, in which the replay
distribution is reshaped during training to track the model's current
weakness rather than mirroring buffer composition; their results on
vision benchmarks suggest that what is replayed matters as much as how
much. Guo and Schwaller (2023), although working in de novo molecular
design rather than security, demonstrate that anchoring replay on a
stable reference distribution accelerates learning and stabilises the
policy across drifting objectives, a principle that translates directly
to anchoring the benign class in an IDS replay buffer. The most recent
continual-learning IDS work reinforces this frontier: Zhang et
al.~(2025) frame malware-traffic classification as expandable
class-incremental learning with neural architecture search; Guo et
al.~(2025a) combine clustering-based memory, multi-level knowledge
distillation, and meta-learning reweighting for intrusion detection
under evolving threats; and Chen et al.~(2025) apply Synaptic
Intelligence regularisation to a convolutional IDS on CICIDS2017 to
counter forgetting under class imbalance. None, however, keeps benign
traffic present in every experience, the exposure pattern our CII
scenario targets. Taken together, these works reinforce the position
that replay is the right family for streaming IDS, but none of them
studies what happens when the dominant class (benign) is forced into a
single early experience, which is precisely the evaluation gap our CII
scenario addresses.

\subsection{Adversarial attacks on learning-based
detectors}\label{adversarial-attacks-on-learning-based-detectors}

A learning-based IDS is a target as well as a defence, and the
adversarial machine learning literature has mapped the ways an attacker
can subvert one. Evasion attacks perturb inputs at test time; poisoning
attacks corrupt the training data itself, a threat first formalised for
support vector machines by Biggio et al.~(2012) and surveyed across a
decade of work by Biggio and Roli (2018). In the network security
setting, Apruzzese et al.~(2021) show that realistic poisoning and
evasion attacks degrade deep NIDS models even under tight attacker
constraints. Continual learning widens this surface in a way static
training does not: the replay buffer is persistent, trusted state that
is consulted at every future update, so a single corruption propagates
forward in time. Two poisoning families are relevant to this paper and
differ in objective and evaluation criterion.

\subsubsection{Label-flipping attacks}\label{label-flipping-attacks}

Label flipping is the simplest instance of poisoning: the attacker
reassigns the class labels of training samples while leaving the feature
values untouched, corrupting the supervision signal that shapes the
decision boundaries. Its objective is overt degradation, so its effect
can be verified directly in the performance metrics of the poisoned
model. In a continual learner the natural target is the replay buffer,
because a flipped exemplar is replayed at every subsequent update and
its damage compounds across the stream. Label flipping is one of the two
attacks quantified in this paper, and its effect on a benign-anchored
class-balanced IDS buffer has not previously been measured.

\subsubsection{Backdoor attacks}\label{backdoor-attacks}

Backdoor attacks are a subcategory of poisoning with a specific trigger
mechanism: the attacker implants a fixed feature pattern into a subset
of poisoned samples labelled as the target class, so that the trained
model misclassifies any input carrying the trigger while behaving
normally on clean inputs. The objective is stealth rather than
degradation, so success is measured by the attack success rate on
triggered inputs, not by aggregate accuracy. In flow-based NIDS the
trigger must live in the derived tabular features rather than in raw
payload; timing statistics such as the packet inter-arrival-time
features are a natural trigger channel, because an attacker controls
them directly by pacing packet transmission and small timing
perturbations leave the volume and flag features untouched. In the
continual setting, Umer et al.~(2020) demonstrate targeted forgetting
and false-memory implantation in continual learners through adversarial
backdoor samples, and Guo et al.~(2024) show that backdoors introduced
during continual learning persist across subsequent tasks even when only
a single task's training is compromised, which makes the persistent
replay buffer an especially attractive implantation site. This paper
quantifies such a backdoor, with a packet inter-arrival-time trigger, on
the replay buffer of a benign-anchored class-balanced IDS.

\subsection{Research gap}\label{research-gap}

\begin{table*}[t]
\rmfamily
\small
\caption{Representative prior work positioned along the four dimensions this paper combines: tabular transformer encoder, continual-learning (CL) strategy, evaluation scenario, and adversarial investigation of the learner. The final row positions this paper.}\label{tbl1}
\begin{tabular*}{\tblwidth}{@{}p{2.4cm}p{2.4cm}Cp{2.4cm}p{1.6cm}p{2.1cm}p{2.6cm}@{}}
\toprule
Work / Method & Approach & Tab.\ transf. & CL strategy & Scenario & Adv.\ inv. & Key limitation \\
\midrule
CNN/RNN IDS (Yin et al., 2017); Autoencoder IDS (Shone et al., 2018) & CNN/RNN and autoencoder detectors on flow records & No & --- & Static & No & Flat-vector or pseudo-image inputs; no label-set change \\[5pt]
TabTransformer (Huang et al., 2020); FT-Transformer/SAINT (Gorishniy et al., 2021; Somepalli et al., 2021); TabNet (Arik \& Pfister, 2021) & Attention-based encoders for tabular data & Yes & --- & Static & No & Stationary benchmarks; no streaming classes or forgetting \\[5pt]
EL-GNN (Nguyen \& Park, 2025) & EWC-style regularisation on a graph-convolutional IDS & No & Regularisation (EWC-style) & Task-incremental & No & Fragile under NIDS class imbalance \\[5pt]
SPIDER (Amalapuram et al., 2024) & Gradient-projection memory; label-efficient ($\sim$20\% labels) & No & Gradient-projection memory & CI & No & No exemplar memory; benign exposure pattern not modelled \\[5pt]
TAMR (A H et al., 2025) & Task-aware replay scored by current-task relevance & No & Task-aware replay & CI & No & Benign appears only in the first experience \\[5pt]
ECBRS / PAPA (Amalapuram et al., 2023) & Class-balanced reservoir sampling; interference approximation & No & Class-balanced reservoir replay & CI & No & Benign concentrated early; no CII evaluation \\[5pt]
Amalapuram et al. (2022); Smith et al. (2024); Guo \& Schwaller (2023) & Early CL-for-NIDS; adaptive and anchored replay & No & Replay variants & CI; non-IDS domains & No & Predates IDS-specific buffer management; includes non-NIDS domains with limited direct applicability \\[5pt]
This paper & Tabular transformer & Yes & Benign-anchored class-balanced replay & CI + CII & Yes (replay-buffer poisoning) & Addresses all above: contextual encoder, benign anchored every experience, production-realistic CII scenario \\
\bottomrule
\end{tabular*}
\end{table*}

The literature therefore converges on three observations: (i) tabular
transformers materially improve representation learning for
heterogeneous flow features compared with MLPs and classical deep
learners (Huang et al., 2020; Gorishniy et al., 2021); (ii) replay-based
continual learning dominates on severely imbalanced intrusion
benchmarks, with regularisation and gradient-projection methods
competitive only in narrower settings; and (iii) existing scenarios do
not capture the continuous presence of benign traffic. The precise gap
this paper fills is the combination of a transformer-based tabular
encoder with a benign-anchored class-balanced replay buffer, evaluated
under a scenario that keeps benign traffic visible at every experience.

\section{Methodology}\label{methodology}

\subsection{Problem formulation}\label{problem-formulation}

Let a network flow record be a feature vector \(x \in \mathbb{R}^d\),
where d = 78 features derived from the CICIDS2017 CICFlowMeter pipeline.
The label set \(C = \{0, 1, \ldots, K-1\}\) is partitioned into a
sequence of experiences \(E_0, E_1, \ldots, E_T\), each associated with
a disjoint subset of attack classes (and, under CII, the benign class
0). At experience E\_k the learner receives a labelled training set
\(D_k\) and must update its classifier \(f_\theta\) without accessing
prior data \(D_0, \ldots, D_{k-1}\). The goal is to minimise forgetting
F and intransigence I simultaneously: a method is operationally viable
only if both are small. Forgetting is defined as the drop from the best
accuracy previously achieved on an earlier task to its accuracy after
the final experience, averaged over all earlier tasks; intransigence is
the gap between the accuracy a joint oracle achieves when trained on the
union of all data and the accuracy the continual learner achieves on the
same evaluation. Both are formalised in Section 3.7.

\subsection{Proposed adaptive IDS
framework}\label{proposed-adaptive-ids-framework}

A network flow is uniquely identified by its five-tuple: source IP
address, destination IP address, source port, destination port, and
transport-layer protocol. The framework converts each flow into a
feature vector using the CICFlowMeter pipeline and classifies it into
one of the attack classes observed so far, enabling incremental
deployment as new threat families emerge. The framework comprises four
components, shown in Fig. 1: a preprocessing stage that standardises
continuous features and computes per-column vocabularies for the
categorical features (Section 4), a tabular transformer encoder whose
classifier head expands as new classes appear (Section 3.3), a
benign-anchored class-balanced replay buffer that is rebuilt at the end
of each experience (Section 3.4), and a per-experience training loop
with early stopping (Section 3.5). Sections 3.6 and 3.7 define the
evaluation scenarios and metrics, and Sections 3.8 and 3.9 introduce the
adversarial threat models under which the buffer itself is attacked.

\begin{figure*}[t]
\centering
\includegraphics[width=0.85\linewidth]{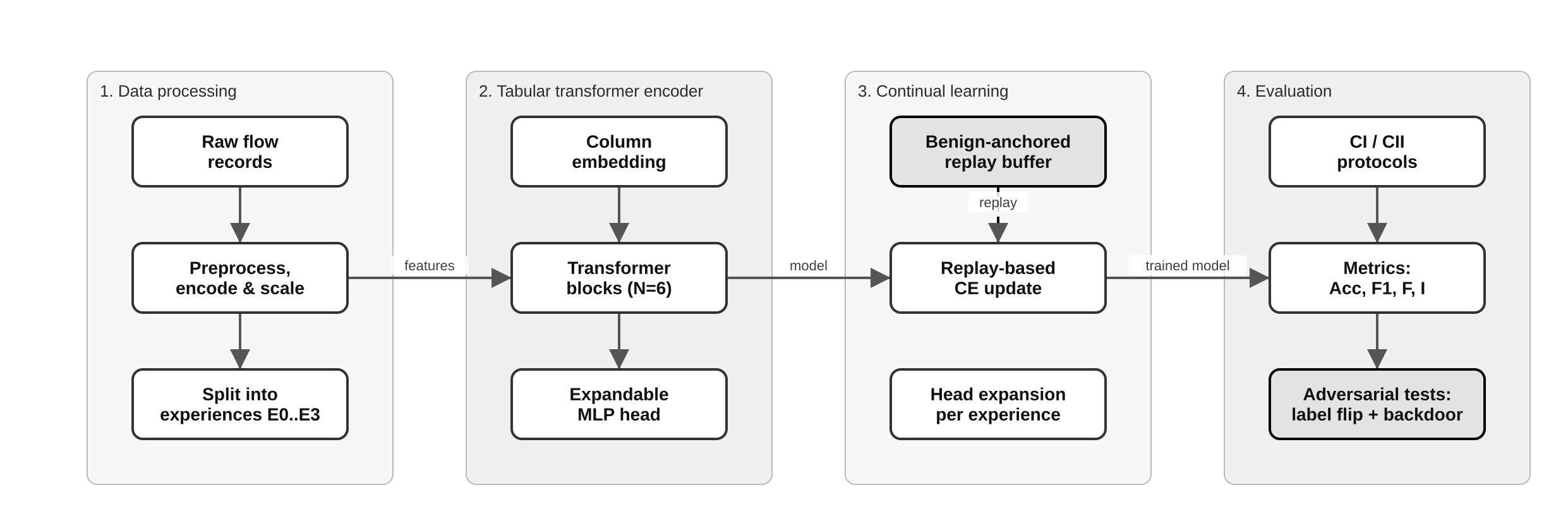}
\caption{Overview of the proposed adaptive IDS framework.}\label{fig1}
\end{figure*}

\subsection{Tabular transformer
encoder}\label{tabular-transformer-encoder}

A flow record is a heterogeneous tuple of numerical descriptors (packet
counts, byte counts, inter-arrival statistics) and categorical
identifiers (protocol, flags, service). The proposed framework embeds
each categorical feature into a learned vector of dimension d\_e,
concatenates it with the numerical features after layer normalisation,
and feeds the resulting token sequence into N stacked transformer
blocks. Each block consists of multi-head self-attention followed by a
position-wise feed-forward network, with residual connections and layer
normalisation around each sub-layer, following the original formulation
of Vaswani et al.~(2017) and the tabular adaptation of Huang et
al.~(2020). Self-attention is the critical ingredient: the
representation of the protocol token is produced by a weighted
combination of all other tokens in the row, so the same port number can
encode different semantics depending on the protocol and flag values
that accompany it. An MLP head with one hidden layer maps the pooled
transformer output to a softmax over the union of classes seen so far.
When new classes appear at an experience boundary, the head is expanded
by appending output units for the new classes, and the weights and
biases of the existing output units are copied over unchanged, so the
decision behaviour learned for earlier classes is preserved at the start
of each new experience.

\begin{figure*}[t]
\centering
\includegraphics[width=0.85\linewidth]{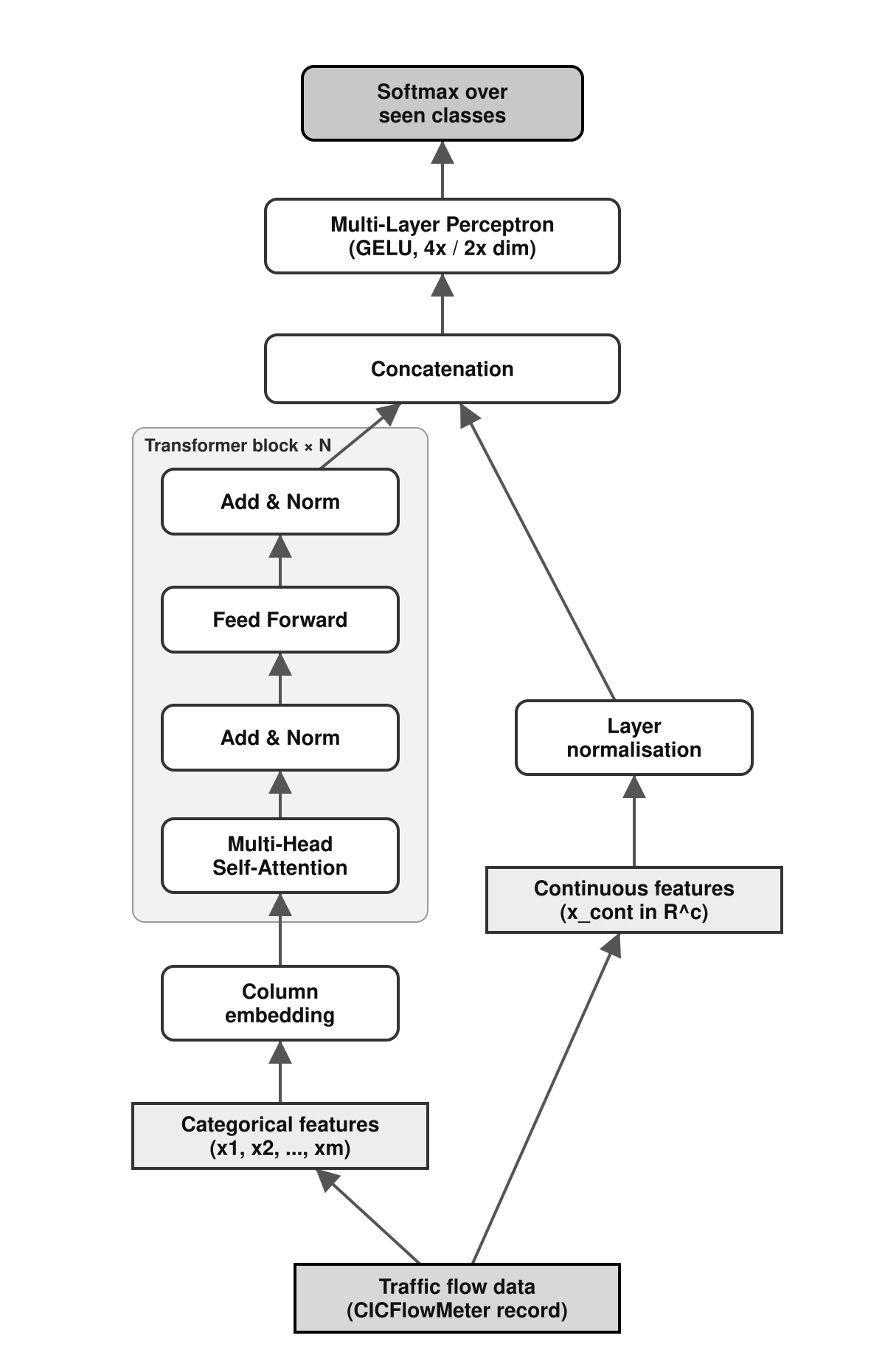}
\caption{Tabular transformer encoder architecture.}\label{fig2}
\end{figure*}

\subsection{Benign-anchored class-balanced experience
replay}\label{benign-anchored-class-balanced-experience-replay}

Let M denote the replay buffer with per-class memory budget b, expressed
as the percentage of each class's training flows retained. At the end of
each experience, the buffer is updated by class-balanced reservoir
sampling (Chrysakis \& Moens, 2020), each class seen so far contributes
the b\% of its training flows that fits the budget, with uniform-random
admission within a class and class-wise random eviction when the
per-class budget is exceeded. During training on E\_k, the replay buffer
M is concatenated with the current experience's training set at the
data-loading stage, and each batch of size B is drawn from this combined
pool. Training minimises categorical cross-entropy, the negative
log-likelihood of the true class under the model's softmax output,
computed over these mixed batches. The benign-anchoring rule is the
operational counterpart of the CII assumption that benign traffic is
always present, and is crucial because the dominant source of confusion
during incremental training is the benign--new-attack decision boundary.
The benign class is therefore guaranteed representation in the replay
buffer at every experience, regardless of the current experience's class
mix. If benign were absent from the replayed data in an experience that
introduces a new attack class, the model could collapse large regions of
the benign manifold into the new attack, inflating false positives.

\begin{figure*}[t]
\centering
\includegraphics[width=0.85\linewidth]{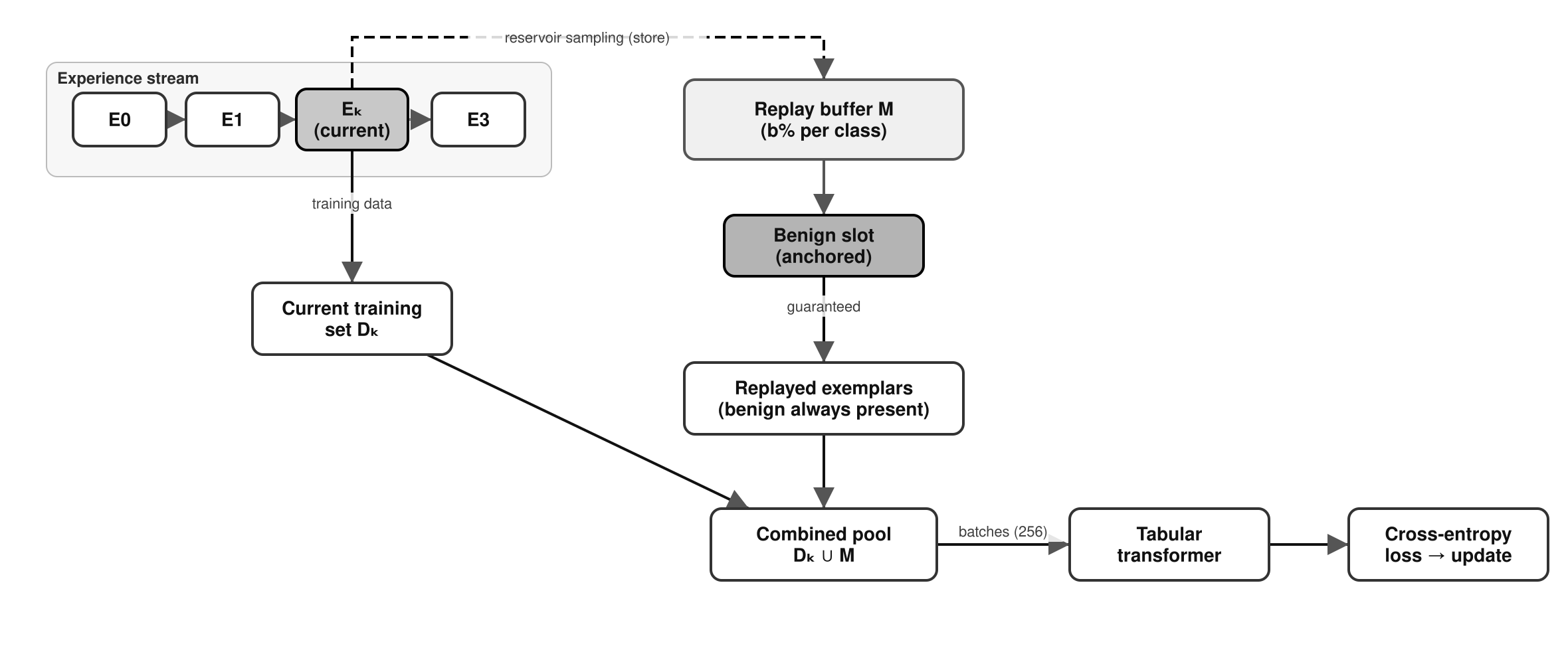}
\caption{Benign-anchored class-balanced experience replay: buffer composition and batch construction.}\label{fig3}
\end{figure*}

\begin{figure*}[t]
\centering
\includegraphics[width=0.85\linewidth]{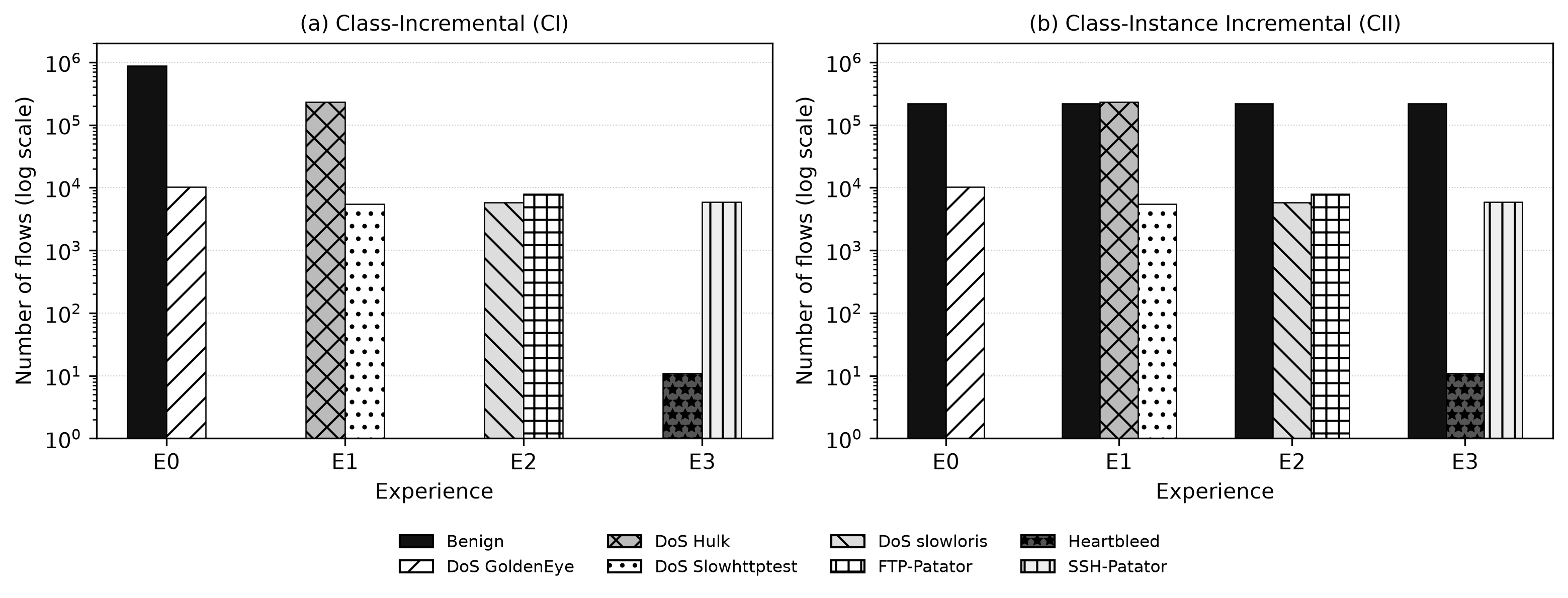}
\caption{Experience distribution under the CI and CII scenarios on CICIDS2017.}\label{fig4}
\end{figure*}

\subsection{Training procedure}\label{training-procedure}

All hyperparameters were selected by grid search on the validation set
of the first experience. The network is trained with the AdamW optimiser
at learning rate 5\(\times 10^{-4}\) and weight decay
1\(\times 10^{-5}\), with early stopping on validation loss (patience 10
epochs, up to 100 epochs per experience) and a ReduceLROnPlateau
schedule (factor 0.1, patience 5). Training uses a batch size of 256,
with each batch drawn from the concatenation of the current experience's
training set and the replay buffer; the replay share of a batch
therefore reflects the buffer's proportion of the combined pool rather
than a fixed ratio. The per-class memory budget b, the percentage of
each class's training flows retained in the buffer, is swept over \{1\%,
5\%, 10\%\} to characterise the memory--performance trade-off; we report
all three. The categorical embedding dimension is 32, the encoder uses N
= 6 transformer blocks with 10 attention heads and 0.1 attention and
feed-forward dropout, and the MLP head applies GELU activations over
hidden layers of 4\(\times\) and 2\(\times\) the embedding dimension.
All runs use a fixed random seed of 42.

\subsection{Evaluation scenarios: CI and
CII}\label{evaluation-scenarios-ci-and-cii}

We evaluate under two scenarios. In the class-incremental (CI) scenario,
the eight CICIDS2017 classes are partitioned into four experiences of
two classes each, and benign appears only in the first experience. This
is the scenario used by the majority of prior continual-learning IDS
work (A H et al., 2025; Amalapuram et al., 2024; Amalapuram et al.,
2023). In the class-instance incremental (CII) scenario proposed here,
benign is present in every experience alongside the attack classes
introduced in that experience. A deployed IDS, whether on a campus edge
or an enterprise data-centre perimeter, never sees a burst of benign
followed by pure attack streams: it sees benign all the time. CII places
this fact inside the evaluation scenario. The consequence for the
learner is subtle but material: in CII the decision boundary between
benign and the newly introduced attack must be re-learned in every
experience, which stresses the replay mechanism more heavily than CI
does.

\subsection{Metrics}\label{metrics}

We report four evaluation quantities at the end of the final experience.
Accuracy and macro-averaged F1 on the concatenated test set of all
classes capture overall detection performance. The retention metrics are
defined on the accuracy matrix: let \(A_{i,j}\) denote the model's
performance (accuracy or macro-F1) on the test set of task j immediately
after training on experience i, for tasks and experiences indexed 1,
\ldots, T with T = 4.

Forgetting, following Chaudhry et al.~(2018), measures how much of the
peak performance on an earlier task has been lost by the end of the
stream. For a single task j it is the drop from the best performance
achieved on that task at any earlier point to its performance after the
final experience, floored at zero:

\begin{equation}
f_j = \max\!\left(0,\ \max_{k \in \{1,\ldots,T-1\}} A_{k,j} - A_{T,j}\right),
\end{equation}

and the reported forgetting is the average over all earlier tasks:

\begin{equation}
F = \frac{1}{T-1}\sum_{j=1}^{T-1} f_j.
\end{equation}

Intransigence, introduced in the Riemannian Walk framework of Chaudhry
et al.~(2018), measures the plasticity gap relative to oracle
performance. We compute it at the stream level as the amount by which
the continual learner falls short of the joint oracle on the same
headline metric:

\begin{equation}
I = A_{\text{joint}} - A_{\text{CL}},
\end{equation}

where A\_joint is the joint oracle's overall accuracy (or macro-F1) and
A\_CL is the continual learner's value of the same quantity; small
negative values indicate that the learner matches or marginally exceeds
the single-snapshot oracle. Reporting forgetting and intransigence
together is essential because a trivially stable model that refuses to
learn has zero forgetting but high intransigence. The joint model is
trained end-to-end on the union of experiences and provides the accuracy
and F1 upper bounds.

\subsection{Adversarial threat model: label flipping on the replay
buffer}\label{adversarial-threat-model-label-flipping-on-the-replay-buffer}

The replay buffer that stabilises the framework is also a new attack
surface: unlike the model weights, the buffer persists between updates
and is consulted at every subsequent experience, so an attacker who
corrupts it once influences all future training. We adopt a
data-poisoning threat model in which the attacker has write access to
the stored exemplars (for example through compromise of the exemplar
store or of the labelling pipeline that feeds it) but no access to the
model parameters or gradients. The attack is label flipping: the label
of every stored exemplar is reassigned to a different class drawn
uniformly from the remaining classes present in the buffer. The flip is
applied after the buffer is rebuilt at the end of each experience, so
the poisoned exemplars are replayed during every later update. The flip
rate on the buffer's stored contents is fixed at 100\%: every exemplar
admitted to the buffer is flipped, for every budget tested. The poison
rate p in this study therefore denotes the per-class buffer budget, not
the flip rate: the percentage of each class's training flows that is
stored, and subsequently flipped in full, in the buffer. It therefore
bounds the volume of poisoned exemplars replayed at each update, rather
than a per-sample flipping probability. We evaluate p in \{1\%, 5\%,
10\%\} under the CI scenario, keeping the seed and training procedure
identical to the corresponding clean runs; the attack implementation is
included in the released code.

\begin{figure*}[t]
\centering
\includegraphics[width=0.85\linewidth]{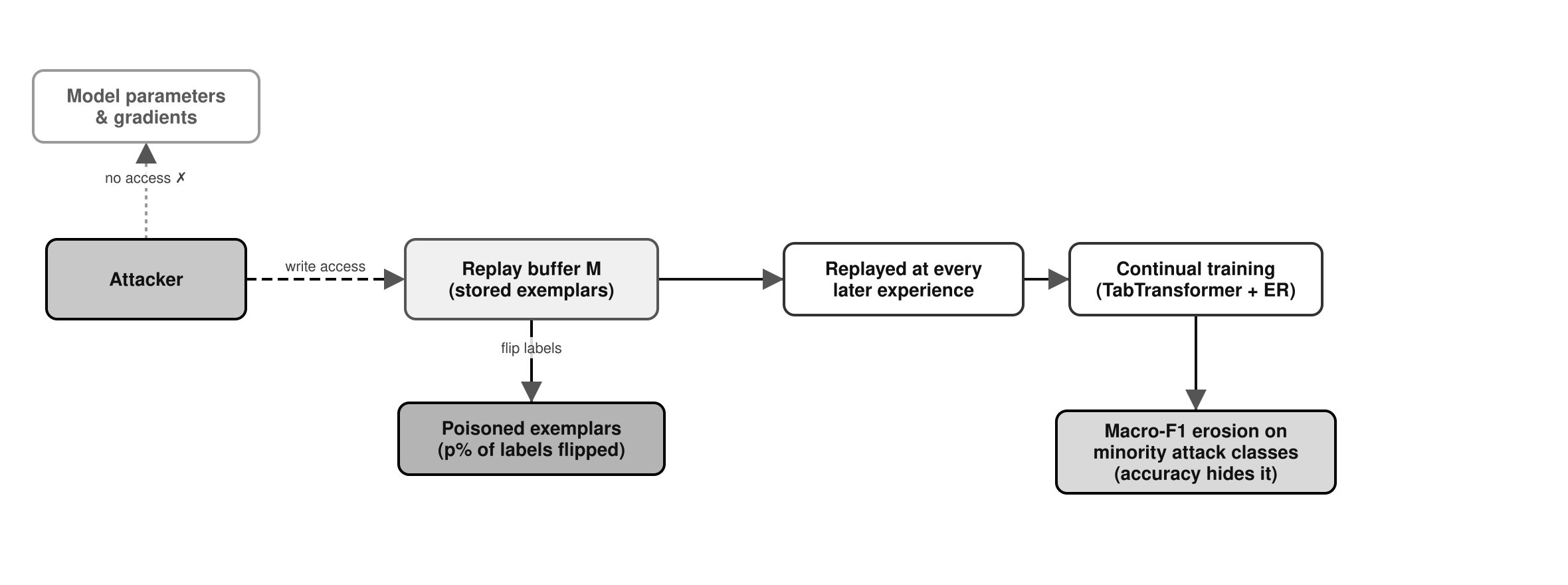}
\caption{Label-flipping poisoning attack on the replay buffer under the adopted threat model.}\label{fig5}
\end{figure*}

\subsection{Adversarial threat model: backdoor attack on the replay
buffer}\label{adversarial-threat-model-backdoor-attack-on-the-replay-buffer}

Label flipping degrades the model openly and is verified directly in the
performance metrics. A backdoor attack on the same surface pursues the
opposite objective: stealth. Under the same attacker capability as
Section 3.8 (write access to the stored exemplars, no access to model
parameters or gradients), the attacker injects a trigger into a subset
of stored exemplars: a fixed perturbation of selected feature values,
with the exemplar labels set to benign. In the flow-feature space of
Section 3.1 the natural trigger channel is the timing statistics, in
particular the packet inter-arrival-time features listed in Table 2,
because an attacker can realise the trigger on live traffic by pacing
packet transmission while leaving volume, flag, and header-derived
features untouched.

\begin{table}[t]
\rmfamily
\caption{CICIDS2017 inter-arrival-time (IAT) features used to construct the backdoor trigger.}\label{tbl2}
\begin{tabular*}{\tblwidth}{@{}L@{}}
\toprule
Feature \\
\midrule
Flow IAT Mean \\
Flow IAT Std \\
Flow IAT Max \\
Flow IAT Min \\
Fwd IAT Total \\
Fwd IAT Mean \\
Fwd IAT Std \\
Fwd IAT Max \\
Fwd IAT Min \\
Bwd IAT Total \\
Bwd IAT Mean \\
Bwd IAT Std \\
Bwd IAT Max \\
Bwd IAT Min \\
\bottomrule
\end{tabular*}
\end{table}

Once the poisoned exemplars are replayed across subsequent experiences,
the model learns to associate the trigger pattern with the benign class,
following the clean-label IAT-trigger construction of Guo et
al.~(2025b); at test time, attack flows carrying the trigger are then
misclassified as benign while performance on clean traffic is
essentially unchanged. The evaluation criterion therefore differs from
Section 3.8: a backdoor is measured by its attack success rate, the
fraction of trigger-carrying flows classified as benign, alongside clean
accuracy to confirm stealth, rather than by aggregate degradation.
Sections 5.5 and 5.6 quantify both threat models under the CI scenario:
label flipping in Section 5.5 and this backdoor, with its attack success
rate, in Section 5.6.

\begin{figure*}[t]
\centering
\includegraphics[width=0.85\linewidth]{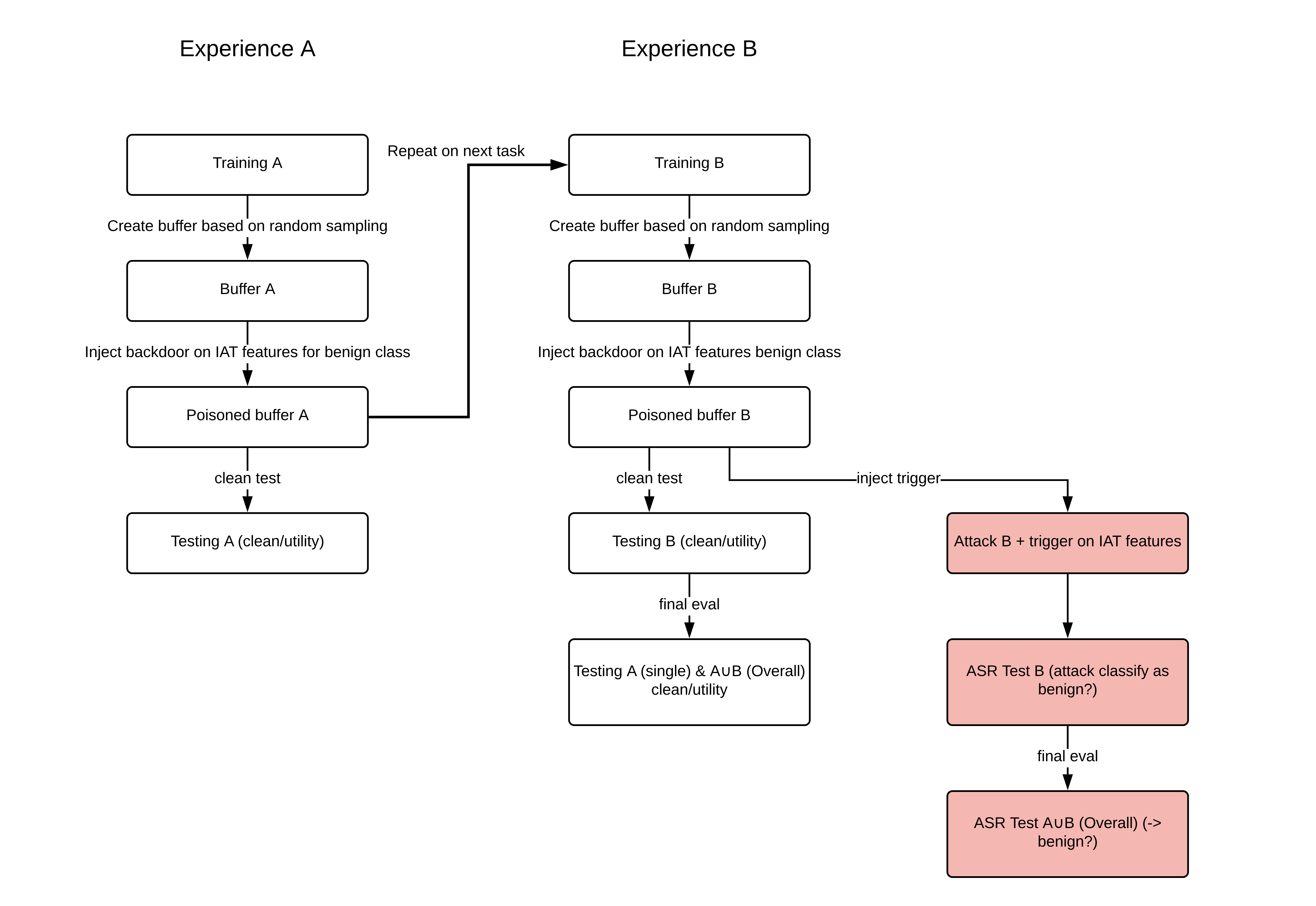}
\caption{Backdoor attack pipeline across two experiences: the buffer is rebuilt by random sampling after each experience, the trigger is injected into the stored inter-arrival-time features with the exemplar labels set to benign, and evaluation splits into clean testing (utility) and attack-success-rate testing (the fraction of triggered flows classified as benign) at both the single-experience and cumulative levels.}\label{fig6}
\end{figure*}

\section{Experimental Setup}\label{experimental-setup}

All experiments use the CICIDS2017 benchmark (Sharafaldin et al., 2018).
After deduplication and removal of invalid flows, the 8-class dataset is
split 60/20/20 into training, validation, and test partitions. Table 3
reports the final class distribution. The distribution is highly
imbalanced: the benign class accounts for roughly 77\% of flows, while
the Heartbleed class accounts for 11 flows out of 1,138,612. This
imbalance, characteristic of real-world network traffic, is the primary
reason why a benign-aware replay policy is necessary.

\begin{table*}[t]
\rmfamily
\caption{CICIDS2017 class distribution after preprocessing.}\label{tbl3}
\begin{tabular*}{\tblwidth}{@{}CLRRRR@{}}
\toprule
Class ID & Class Name & Train & Validation & Test & Total \\
\midrule
0 & Benign & 523,263 & 174,421 & 174,421 & 872,105 \\
1 & DoS GoldenEye & 6,176 & 2,058 & 2,059 & 10,293 \\
2 & DoS Hulk & 138,644 & 46,214 & 46,215 & 231,073 \\
3 & DoS Slowhttptest & 3,299 & 1,100 & 1,100 & 5,499 \\
4 & DoS slowloris & 3,477 & 1,160 & 1,159 & 5,796 \\
5 & FTP-Patator & 4,763 & 1,587 & 1,588 & 7,938 \\
6 & Heartbleed & 7 & 2 & 2 & 11 \\
7 & SSH-Patator & 3,538 & 1,180 & 1,179 & 5,897 \\
\midrule
 & Total & 683,167 & 227,722 & 227,723 & 1,138,612 \\
\bottomrule
\end{tabular*}
\end{table*}

Preprocessing replicates the pipeline of Sharafaldin et al.~(2018):
infinite and NaN values are dropped, categorical features retain their
original values, with a per-column vocabulary size computed and passed
to the embedding layers, and continuous features are scaled to unit
variance with statistics computed on the training partition only. Table
4 describes the experience partitioning for both scenarios.

\begin{table*}[t]
\rmfamily
\caption{Experience partitioning for the CI and CII scenarios on CICIDS2017.}\label{tbl4}
\begin{tabular*}{\tblwidth}{@{}p{1.2cm}p{6.6cm}p{8.6cm}@{}}
\toprule
Exp. & CI scenario (classes) & CII scenario (classes) \\
\midrule
$E_0$ & 0 (Benign), 1 (DoS GoldenEye) & 0 (Benign), 1 (DoS GoldenEye) \\
$E_1$ & 2 (DoS Hulk), 3 (DoS Slowhttptest) & 0 (Benign), 2 (DoS Hulk), 3 (DoS Slowhttptest) \\
$E_2$ & 4 (DoS slowloris), 5 (FTP-Patator) & 0 (Benign), 4 (DoS slowloris), 5 (FTP-Patator) \\
$E_3$ & 6 (Heartbleed), 7 (SSH-Patator) & 0 (Benign), 6 (Heartbleed), 7 (SSH-Patator) \\
\bottomrule
\end{tabular*}
\end{table*}

We compare seven continual-learning strategies: (i) Na"ive sequential
fine-tuning as a lower bound; (ii) elastic weight consolidation (EWC)
(Kirkpatrick et al., 2017), which adds a quadratic penalty on parameters
deemed important for earlier tasks; (iii) learning without forgetting
(LwF) (Li \& Hoiem, 2018), a distillation-based baseline that constrains
outputs on new data to match earlier predictions; (iv) iCaRL (Rebuffi et
al., 2017), which combines herding-based exemplar selection with a
nearest-mean-of-exemplars classifier; (v) ER-Stratified, experience
replay whose buffer samples each class in proportion to its training
frequency (Chaudhry et al., 2019b), included under CII to isolate the
effect of benign anchoring; (vi) ER-Balanced, experience replay with
class-balanced buffer management (Chrysakis \& Moens, 2020), which
coincides with the proposed framework in the CII scenario; and (vii) the
joint (oracle) model trained on the union of all experiences, which
provides an upper bound on accuracy and F1. All strategies share the
TabTransformer backbone of Section 3.3 and the training configuration of
Table 5; they differ only in how knowledge from earlier experiences is
preserved. Strategy-specific hyperparameters follow common practice: EWC
uses regularisation strength \(\lambda = 1000\) with the Fisher
information estimated on 4,096 samples, LwF uses distillation weight
\(\alpha = 0.5\) and temperature \(T = 2\), and iCaRL uses a fixed total
exemplar budget of 2,000. Table 5 lists the hyperparameter configuration
used across all methods.

\begin{table*}[t]
\rmfamily
\caption{Hyperparameter configuration.}\label{tbl5}
\begin{tabular*}{\tblwidth}{@{}p{3.3cm}p{4.5cm}p{3.5cm}p{4.5cm}@{}}
\toprule
Hyperparameter & Value & Hyperparameter & Value \\
\midrule
Optimiser & AdamW & Transformer blocks $N$ & 6 \\
Learning rate & $5{\times}10^{-4}$ & Attention heads & 10 \\
Weight decay & $1{\times}10^{-5}$ & Categorical embedding dim & 32 \\
Batch size & 256 & MLP head hidden dim & $4\times$ and $2\times$ embedding dim (GELU) \\
LR scheduler & ReduceLROnPlateau on validation loss (factor 0.1, patience 5) & Per-class memory budget $b$ (\%) & \{1\%, 5\%, 10\%\} \\
Loss function & Categorical cross-entropy & Benign anchored in replay & Yes \\
Epochs per experience & up to 100 (early stopping, patience 10) & Random seed & 42 \\
\bottomrule
\end{tabular*}
\end{table*}

Experiments are run on a single NVIDIA RTX A5000 GPU (24 GB). Code for
full reproduction is available at \url{https://github.com/um-csnet/ReplayIDS}.

\section{Results and Analysis}\label{results-and-analysis}

\subsection{Scenario 1: class-incremental
(CI)}\label{scenario-1-class-incremental-ci}

Table 6 reports overall accuracy, macro-F1, average forgetting, and
intransigence at the end of the final experience under the CI scenario.
The joint oracle achieves 0.9997 accuracy and 0.9979 macro-F1,
confirming that the dataset is learnable when all classes are available
simultaneously. Sequential fine-tuning (Na"ive) collapses to 0.0052
accuracy and 0.0014 macro-F1 after four experiences: having learned only
the final experience's attack classes, the model classifies essentially
every flow as belonging to those classes and retains nothing from
earlier ones. EWC (Kirkpatrick et al., 2017) does not recover from this
collapse, reaching 0.0324 accuracy and 0.0353 macro-F1; its quadratic
penalty on parameter importance is insufficient to prevent catastrophic
forgetting when the softmax head must accommodate entirely new output
classes at each experience. Both regularisation-based methods are known
to underperform replay in disjoint class-incremental settings (Li \&
Hoiem, 2018; De Lange et al., 2022), and the present results reproduce
this pattern. LwF (Li \& Hoiem, 2018) likewise fails under CI, reaching
0.0699 accuracy and 0.1231 macro-F1: distillation preserves soft-target
consistency with an earlier snapshot but cannot prevent the head from
re-allocating probability mass as the class set expands. iCaRL (Rebuffi
et al., 2017) is the strongest non-replay baseline under CI, reaching
0.8770 accuracy and 0.5448 macro-F1: its nearest-mean-of-exemplars
classifier retains structural class representations even as the encoder
adapts, yielding substantially better retention than
regularisation-based methods, though it falls well short of the oracle.

Experience replay closes the remaining gap to the oracle. ER-Stratified,
included here to give the CI scenario the same replay comparison as
Table 7's CII results, reaches 0.9967 accuracy and 0.9468 macro-F1 at
\(b = 1\%\), rising to 0.9989/0.9893 at b = 5\% and 0.9994/0.9953 at b =
10\%. With a per-class budget of \(b = 1\%\), ER-Balanced reaches 0.9949
accuracy and 0.9352 macro-F1. Increasing the budget to b = 5\% raises
accuracy to 0.9987, and b = 10\% achieves 0.9994 accuracy and 0.9953
macro-F1, within 0.0003 of the oracle on accuracy and 0.0026 on
macro-F1. Forgetting on accuracy decreases monotonically with budget for
both replay variants, from 0.0014 (ER-Stratified) and 0.0020
(ER-Balanced) at \(b = 1\%\) to 0.0003 for both at b = 10\%;
intransigence on accuracy remains at or below 0.0048 throughout. The
best single configuration under CI is tied between ER-Stratified and
ER-Balanced at b = 10\%, which are simultaneously the most stable
(forgetting accuracy = 0.0003) and the most plastic (intransigence
accuracy = 0.0003).

\begin{table*}[t]
\rmfamily
\small
\caption{CI scenario: overall performance after the final experience on CICIDS2017. All methods share the TabTransformer backbone (Section~3.3). Accuracy and macro-F1 are measured on the concatenated test set of all eight classes. Forgetting and intransigence are averaged over earlier tasks. ForgAcc: forgetting on accuracy; IntrAcc: intransigence on accuracy; ForgF1/IntrF1: same for macro-F1. iCaRL uses a fixed total exemplar budget of 2,000 rather than a per-class percentage.}\label{tbl6}
\begin{tabular*}{\tblwidth}{@{}LCCCCCCC@{}}
\toprule
Method & Buffer $b$ (\%) & Accuracy & Macro-F1 & Forg.\ (Acc) & Forg.\ (F1) & Intr.\ (Acc) & Intr.\ (F1) \\
\midrule
Joint (Oracle) & --- & 0.9997 & 0.9979 & --- & --- & --- & --- \\
Na\"ive (Sequential, failed) & --- & 0.0052 & 0.0014 & 0.9998 & 0.9961 & 0.9945 & 0.9965 \\
EWC & --- & 0.0324 & 0.0353 & 0.9112 & 0.9264 & 0.9673 & 0.9626 \\
LwF & --- & 0.0699 & 0.1231 & 0.8122 & 0.8501 & 0.9298 & 0.8748 \\
iCaRL & 2,000 exemplars & 0.8770 & 0.5448 & 0.0534 & 0.0388 & 0.1227 & 0.4531 \\
\multirow{3}{*}{ER-Stratified} & 1 & 0.9967 & 0.9468 & 0.0014 & 0.0064 & 0.0030 & 0.0511 \\
 & 5 & 0.9989 & 0.9893 & 0.0006 & 0.0074 & 0.0008 & 0.0086 \\
 & 10 & 0.9994 & 0.9953 & 0.0003 & 0.0059 & 0.0003 & 0.0026 \\
\multirow{3}{*}{ER-Balanced (ours)} & 1 & 0.9949 & 0.9352 & 0.0020 & 0.0176 & 0.0048 & 0.0627 \\
 & 5 & 0.9987 & 0.9283 & 0.0005 & 0.0063 & 0.0010 & 0.0696 \\
 & 10 & 0.9994 & 0.9953 & 0.0003 & 0.0059 & 0.0003 & 0.0026 \\
\bottomrule
\end{tabular*}
\end{table*}

\begin{figure*}[t]
\centering
\includegraphics[width=0.85\linewidth]{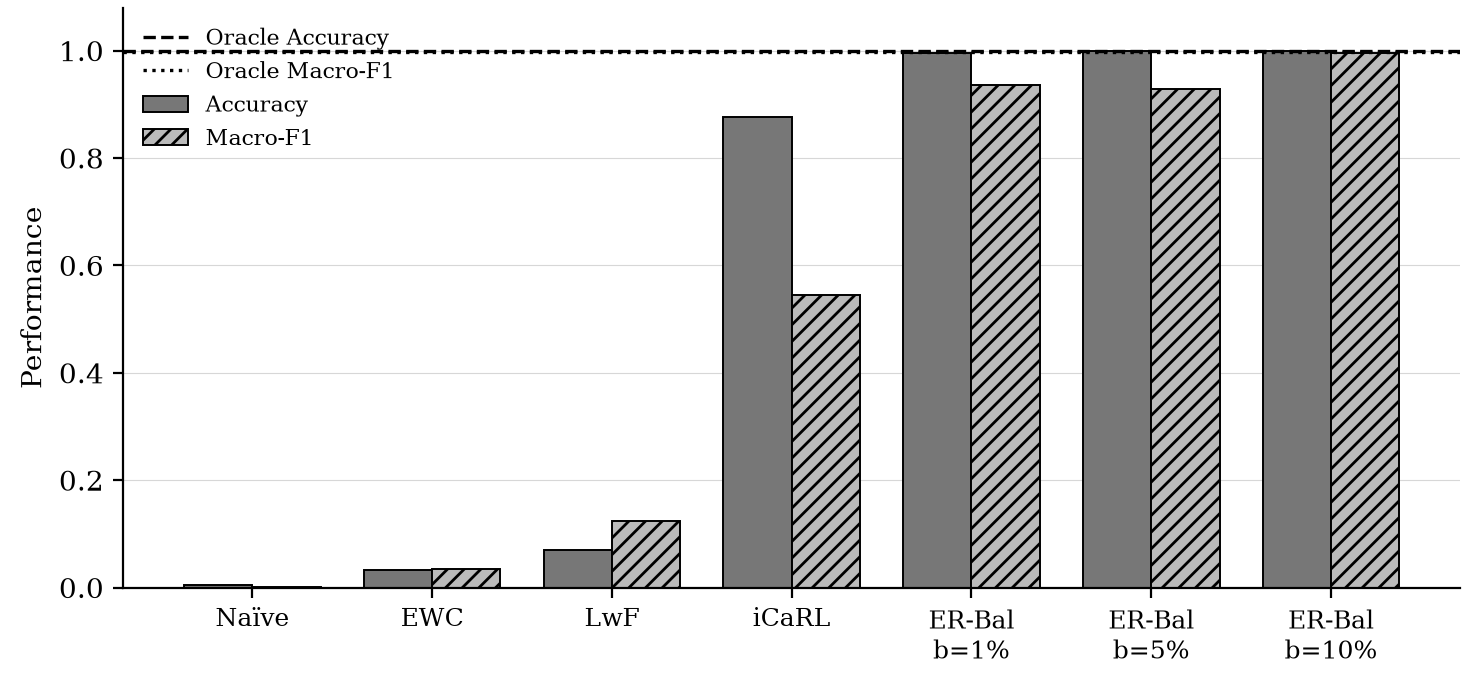}
\caption{CI scenario: accuracy and macro-F1 by continual-learning method. Dashed lines indicate oracle accuracy (0.9997) and macro-F1 (0.9979). ER-Balanced at $b=10\%$ closes within 0.0003 of the oracle on accuracy.}\label{fig7}
\end{figure*}

\subsection{Scenario 2: class-instance incremental
(CII)}\label{scenario-2-class-instance-incremental-cii}

Table 7 reports performance under the CII scenario, in which benign
traffic is present in every experience. Here accuracy and macro-F1 are
averaged over the four experience checkpoints, so they summarise
behaviour across the whole stream; Table 8 complements this view with
overall performance at the final experience only, mirroring the CI
scenario of Table 6. The joint oracle remains at 0.9955 accuracy and
0.9795 macro-F1 under CII. The Na"ive sequential model reaches 0.8700
checkpoint-averaged accuracy and 0.6278 macro-F1: continuous benign
exposure partially buffers the model against catastrophic forgetting,
yet macro-F1 below 0.63 indicates that attack-class knowledge is still
substantially lost. Under CII, the regularisation-based methods recover
partially: EWC reaches 0.8698 accuracy and 0.6295 macro-F1, and LwF
reaches 0.9096 accuracy and 0.7255 macro-F1, both well above their
near-zero CI values. iCaRL improves further to 0.9546 accuracy and
0.8276 macro-F1. The continuous presence of benign traffic prevents the
complete head-collapse seen under CI for all non-rehearsal strategies,
but their macro-F1 remains below 0.83, indicating that attack-class
discrimination is still impaired relative to the oracle. ER-Stratified
with b = 10\% reaches 0.9987 accuracy and 0.9945 macro-F1, and
ER-Balanced with b = 10\% achieves 0.9989 accuracy and 0.9962 macro-F1,
both marginally exceeding the oracle ceiling on accuracy (negative
intransigence), reflecting the advantage of the class-balanced rehearsal
schedule over a single-snapshot oracle.

\begin{table*}[t]
\rmfamily
\small
\caption{CII scenario: performance across the learning stream on CICIDS2017. All methods share the TabTransformer backbone (Section~3.3). Accuracy and macro-F1 are the mean of the cumulative overall accuracy and macro-F1 measured after each of the four experiences. Small negative intransigence values indicate that the continual learner matches or slightly exceeds the oracle, and reflect seed variation rather than super-oracle performance. iCaRL uses a fixed total exemplar budget of 2,000 rather than a per-class percentage.}\label{tbl7}
\begin{tabular*}{\tblwidth}{@{}LCCCCCCC@{}}
\toprule
Method & Buffer $b$ (\%) & Accuracy (mean) & Macro-F1 (mean) & Forg.\ (Acc) & Forg.\ (F1) & Intr.\ (Acc) & Intr.\ (F1) \\
\midrule
Joint (Oracle) & --- & 0.9955 & 0.9795 & --- & --- & --- & --- \\
Na\"ive (Sequential) & --- & 0.8700 & 0.6278 & 0.1719 & 0.5995 & 0.1255 & 0.3517 \\
EWC & --- & 0.8698 & 0.6295 & 0.1727 & 0.5998 & 0.1257 & 0.3500 \\
LwF & --- & 0.9096 & 0.7255 & 0.1153 & 0.4689 & 0.0859 & 0.2540 \\
iCaRL & 2,000 exemplars & 0.9546 & 0.8276 & 0.0005 & 0.0000 & 0.0409 & 0.1519 \\
\multirow{3}{*}{ER-Stratified} & 1 & 0.9974 & 0.9847 & 0.0022 & 0.0115 & $-$0.0019 & $-$0.0052 \\
 & 5 & 0.9985 & 0.9939 & 0.0016 & 0.0077 & $-$0.0030 & $-$0.0144 \\
 & 10 & 0.9987 & 0.9945 & 0.0005 & 0.0019 & $-$0.0031 & $-$0.0150 \\
\multirow{3}{*}{ER-Balanced (ours)} & 1 & 0.9984 & 0.9879 & 0.0011 & 0.0024 & $-$0.0029 & $-$0.0084 \\
 & 5 & 0.9986 & 0.9936 & 0.0007 & 0.0036 & $-$0.0031 & $-$0.0141 \\
 & 10 & 0.9989 & 0.9962 & 0.0003 & 0.0006 & $-$0.0034 & $-$0.0167 \\
\bottomrule
\end{tabular*}
\end{table*}

\begin{figure*}[t]
\centering
\includegraphics[width=0.85\linewidth]{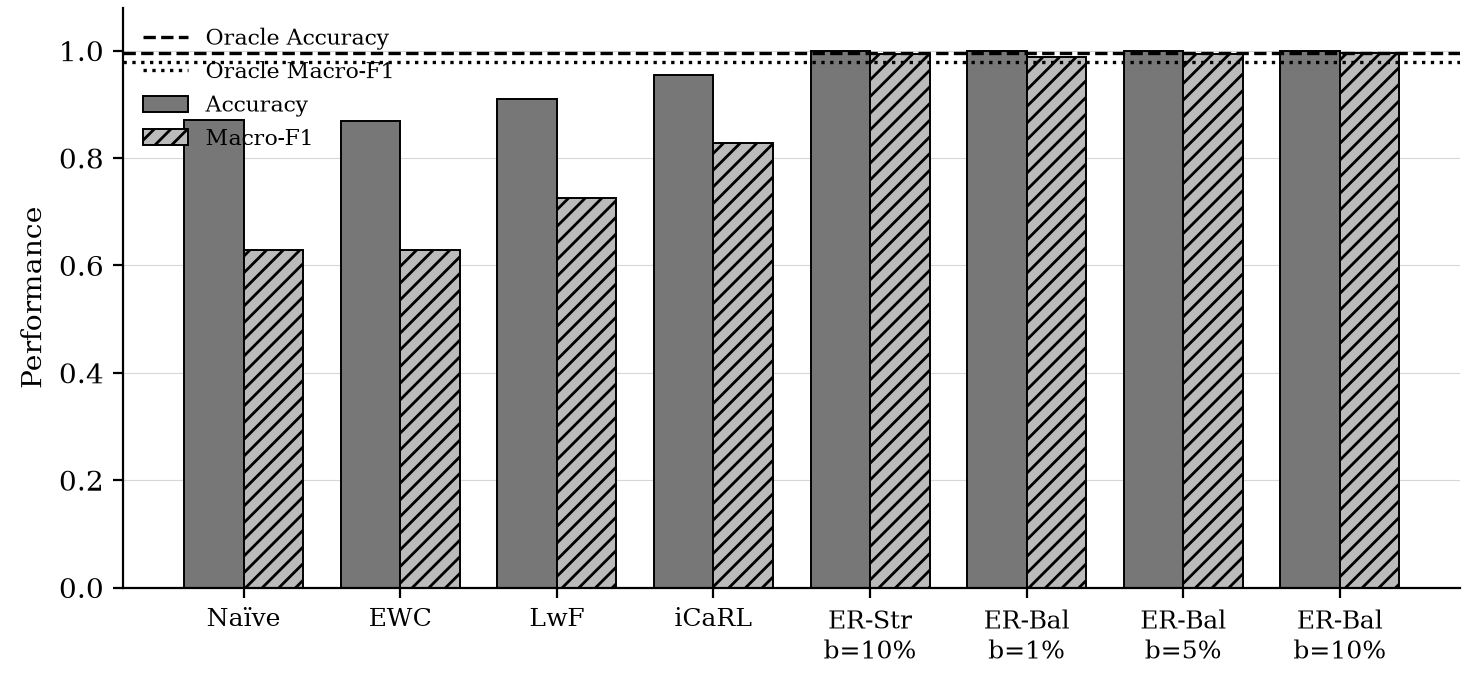}
\caption{CII scenario: accuracy and macro-F1 by continual-learning method. Dashed lines indicate oracle accuracy (0.9955) and macro-F1 (0.9795). ER-Balanced at $b=10\%$ matches or marginally exceeds the oracle on both metrics.}\label{fig8}
\end{figure*}

Because the checkpoint-averaged view in Table 7 blends early
experiences, where few classes have been seen and high accuracy is easy,
with later ones, it flatters methods that degrade over time. Table 8
therefore reports overall accuracy and macro-F1 at the final experience
only, measured on the concatenated test set of all eight classes,
mirroring the CI scenario of Table 6. The endpoint view sharpens the
separation between method families: Na"ive falls to 0.7707 accuracy and
0.3538 macro-F1, EWC tracks it closely at 0.7702 and 0.3486, LwF reaches
0.7827 accuracy and 0.5832 macro-F1, and iCaRL drops to 0.9469 accuracy
and 0.7238 macro-F1. Both replay variants remain near the oracle at
every budget, and ER-Balanced at b = 10\% ends the stream at 0.9989
accuracy and 0.9960 macro-F1, essentially matching the joint oracle.

\begin{table}[t]
\rmfamily
\caption{CII scenario: overall performance at the final experience on CICIDS2017, measured on the concatenated test set of all eight classes (the endpoint counterpart of the checkpoint-averaged view in Table~7). The joint oracle is a single model evaluated once on the full test set. iCaRL uses a fixed total exemplar budget of 2,000 rather than a per-class percentage.}\label{tbl8}
\begin{tabular*}{\tblwidth}{@{}LCCC@{}}
\toprule
Method & Buffer $b$ (\%) & Accuracy & Macro-F1 \\
\midrule
Joint (Oracle) & --- & 0.9955 & 0.9795 \\
Na\"ive (Sequential) & --- & 0.7707 & 0.3538 \\
EWC & --- & 0.7702 & 0.3486 \\
LwF & --- & 0.7827 & 0.5832 \\
iCaRL & 2,000 exemplars & 0.9469 & 0.7238 \\
ER-Stratified & 1 & 0.9965 & 0.9721 \\
ER-Stratified & 5 & 0.9986 & 0.9947 \\
ER-Stratified & 10 & 0.9983 & 0.9906 \\
ER-Balanced (ours) & 1 & 0.9980 & 0.9643 \\
ER-Balanced (ours) & 5 & 0.9980 & 0.9866 \\
ER-Balanced (ours) & 10 & 0.9989 & 0.9960 \\
\bottomrule
\end{tabular*}
\end{table}

\subsection{Discussion of the CI--CII
gap}\label{discussion-of-the-cicii-gap}

Three observations deserve emphasis. First, the Na"ive baseline degrades
far less catastrophically under CII than under CI because benign
exposure in every experience acts as a crude implicit anchor; even so,
macro-F1 falls to 0.6278, which is the number a practitioner should care
about because it weights minority attack classes equally with the benign
majority. Second, the regularisation-based methods (EWC and LwF) recover
substantially under CII --- EWC rises from 0.0324 to 0.8698 accuracy and
LwF from 0.0699 to 0.9096 --- but their macro-F1 remains below 0.73,
indicating that the accuracy gain is driven largely by correct benign
classification rather than improved attack detection. iCaRL shows a
similar pattern but with a stronger underlying class-representation
mechanism, improving from 0.8770 to 0.9546 accuracy and reaching 0.8276
macro-F1 under CII. Third, the replay strategy's benign-anchoring
guarantee keeps ER-Balanced near the oracle under both scenarios; under
CII, ER-Stratified (which samples proportionally and therefore
under-represents benign within the replay buffer for later experiences)
shows slightly higher forgetting at small buffer sizes than ER-Balanced
(0.0022 vs 0.0011 at \(b = 1\%\)). Under CI, where benign is
concentrated in a single early experience for both strategies, this
pattern reverses at small budgets (ER-Stratified forgetting 0.0014 vs
ER-Balanced 0.0020 at \(b = 1\%\); see Table 6), and the two strategies
converge at b = 10\%, indicating that the benign-anchoring advantage is
specific to the continuous benign exposure of CII rather than a
universal property of ER-Balanced.

\subsection{Forgetting and intransigence across
methods}\label{forgetting-and-intransigence-across-methods}

Table 9 summarises forgetting and intransigence jointly across all
methods and scenarios, which is the right view for interpreting the
stability-plasticity trade-off for a deployed IDS. A method is
operationally viable only if both quantities are small. Na"ive lies at
the extreme forgetting corner in both scenarios. EWC and LwF inherit
nearly the same forgetting as Na"ive under CI despite their
regularisation terms, and recover partially under CII but with residual
forgetting that remains unacceptable for a production detector. iCaRL
achieves moderate forgetting under CI and near-zero forgetting under
CII, confirming that exemplar-based class representations provide more
robust retention than regularisation alone. ER-Balanced at b = 10\% is
the only method that is simultaneously near zero on both forgetting and
intransigence in both scenarios, which is the behaviour required if the
detector is to be trusted in production.

\begin{table*}[t]
\rmfamily
\caption{Forgetting and intransigence values per method per scenario, final experience.}\label{tbl9}
\begin{tabular*}{\tblwidth}{@{}LLCCCCC@{}}
\toprule
Scenario & Method & Buffer $b$ (\%) & Forg.\ (Acc) & Forg.\ (F1) & Intr.\ (Acc) & Intr.\ (F1) \\
\midrule
CI & Na\"ive & --- & 0.9998 & 0.9961 & 0.9945 & 0.9965 \\
CI & EWC & --- & 0.9112 & 0.9264 & 0.9673 & 0.9626 \\
CI & iCaRL & --- & 0.0534 & 0.0388 & 0.1227 & 0.4531 \\
CI & LwF & --- & 0.8122 & 0.8501 & 0.9298 & 0.8748 \\
CI & ER-Balanced (ours) & 10 & 0.0003 & 0.0059 & 0.0003 & 0.0026 \\
CII & Na\"ive & --- & 0.1719 & 0.5995 & 0.1255 & 0.3517 \\
CII & EWC & --- & 0.1727 & 0.5998 & 0.1257 & 0.3500 \\
CII & LwF & --- & 0.1153 & 0.4689 & 0.0859 & 0.2540 \\
CII & iCaRL & --- & 0.0005 & 0.0000 & 0.0409 & 0.1519 \\
CII & ER-Stratified & 10 & 0.0005 & 0.0019 & $-$0.0031 & $-$0.0150 \\
CII & ER-Balanced (ours) & 10 & 0.0003 & 0.0006 & $-$0.0034 & $-$0.0167 \\
\bottomrule
\end{tabular*}
\end{table*}

\subsection{Adversarial investigation: label flipping on the replay
buffer}\label{adversarial-investigation-label-flipping-on-the-replay-buffer}

The two poisoning families of Sections 3.8 and 3.9 differ in objective,
and the results bear this out sharply: label flipping is catastrophic
and overt, whereas the backdoor of Section 5.6 is stealthy. Table 10
reports overall performance at the end of the final experience when the
replay buffer is poisoned by label flipping at buffer budgets p of 1\%,
5\%, and 10\%, with 100\% of the buffer's stored exemplars flipped at
every budget, alongside the clean ER configuration for reference. The
result is a near-total collapse that is insensitive to the budget: even
at p = 1\% overall accuracy falls from 0.9995 to 0.0053 and macro-F1
from 0.9957 to 0.2074, and raising the budget to 10\% changes little
(0.0052 accuracy, 0.2077 macro-F1). Forgetting on accuracy is near total
at every budget (0.9987 to 0.9992), meaning the flipped exemplars
overwrite essentially all previously learned decision boundaries, and
intransigence on accuracy near 0.9945 shows the poisoned model can no
longer fit even the current experience. The mechanism is direct: because
the entire buffer is relabelled to wrong classes and replayed at every
subsequent update, the replay signal contradicts the current-task
supervision at every step, and the model converges to a degenerate
solution that classifies almost everything as a single class. Label
flipping is therefore trivial for an attacker to weaponise but equally
trivial for a defender to notice, because the collapse is plainly
visible in the top-line accuracy that any monitoring dashboard tracks.
Its danger lies in denial of service rather than stealth: a single
corruption of the exemplar store disables the detector outright.

\begin{table*}[t]
\rmfamily
\caption{CI scenario under a label-flipping attack on the replay buffer (ER): overall performance after the final experience on CICIDS2017 as a function of the poisoned buffer budget $p$. Forgetting and intransigence follow the definitions of Section~3.7; forgetting is averaged over the three earlier tasks (excluding the task just trained).}\label{tbl10}
\begin{tabular*}{\tblwidth}{@{}LCCCCCC@{}}
\toprule
Poison rate $p$ & Accuracy & Macro-F1 & Forg.\ (Acc) & Forg.\ (F1) & Intr.\ (Acc) & Intr.\ (F1) \\
\midrule
0\% (clean ER) & 0.9995 & 0.9957 & 0.0007 & 0.0025 & 0.0003 & 0.0022 \\
1\% & 0.0053 & 0.2074 & 0.9987 & 0.9963 & 0.9945 & 0.7904 \\
5\% & 0.0052 & 0.2244 & 0.9990 & 0.9966 & 0.9946 & 0.7734 \\
10\% & 0.0052 & 0.2077 & 0.9992 & 0.9973 & 0.9946 & 0.7902 \\
\bottomrule
\end{tabular*}
\end{table*}

\subsection{Adversarial investigation: backdoor attack on the replay
buffer}\label{adversarial-investigation-backdoor-attack-on-the-replay-buffer}

The backdoor attack of Section 3.9 inverts the label-flipping objective:
instead of degrading the model openly, it implants a hidden
trigger-to-benign rule while leaving aggregate performance almost
intact. Table 11 reports overall performance under the backdoor at the
same buffer budgets, and the contrast with Table 10 is stark. Overall
accuracy stays above 0.97 at every budget (0.9992 at p = 1\%, 0.9776 at
p = 10\%), so a top-line accuracy dashboard registers almost nothing.
Macro-F1 is the only aggregate metric that moves, falling from 0.9957
clean to 0.9679 at p = 1\% and 0.8933 at p = 10\%, because overwriting
the timing features of a fraction of the benign exemplars slightly blurs
the benign-attack boundary; even this erosion is modest and easy to
dismiss as noise. The aggregate stealth is exactly the point, and it is
why the backdoor cannot be judged by accuracy or macro-F1 at all.

\begin{table*}[t]
\rmfamily
\caption{CI scenario under a backdoor attack on the replay buffer (ER): overall performance after the final experience on CICIDS2017 as a function of the poisoned buffer budget $p$. Overall accuracy is preserved by design; the attack's effect is captured by the attack success rate in Table~12, not by these aggregate metrics.}\label{tbl11}
\begin{tabular*}{\tblwidth}{@{}LCCCCCC@{}}
\toprule
Poison rate $p$ & Accuracy & Macro-F1 & Forg.\ (Acc) & Forg.\ (F1) & Intr.\ (Acc) & Intr.\ (F1) \\
\midrule
0\% (clean ER) & 0.9995 & 0.9957 & 0.0007 & 0.0025 & 0.0003 & 0.0022 \\
1\% & 0.9992 & 0.9679 & 0.0020 & 0.0054 & 0.0005 & 0.0299 \\
5\% & 0.9982 & 0.9047 & 0.0013 & 0.0127 & 0.0015 & 0.0931 \\
10\% & 0.9776 & 0.8933 & 0.0195 & 0.0216 & 0.0221 & 0.1046 \\
\bottomrule
\end{tabular*}
\end{table*}

The attack's true effect surfaces only in the attack success rate (ASR),
the percentage of the target class's test samples misclassified as
benign, reported in Table 12. Once the trigger is learned, the ASR is
near-total at most budgets and experiences: attack flows that carry the
inter-arrival-time trigger pass the detector as benign essentially every
time, even though the same model still classifies clean traffic at above
0.97 accuracy. This is the defender's worst case, an attacker who can
walk chosen attack traffic straight past a detector whose every top-line
metric looks healthy, and it is precisely the failure mode that the
aggregate view of Table 11 conceals.

\begin{table*}[t]
\rmfamily
\caption{Attack success rate (ASR) of the backdoor on the replay buffer, per poison budget and per experience, under the CI scenario, recomputed from the released training logs. Each cell reports the target class chosen at that experience (selected uniformly at random from classes seen so far) alongside the ASR estimate. Values marked \textdagger\ exceed the theoretical 100\% bound; see the discussion in the text for why.}\label{tbl12}
\begin{tabular*}{\tblwidth}{@{}LLLL@{}}
\toprule
Poison rate $p$ & Experience 1 & Experience 2 & Experience 3 \\
\midrule
1\% & DoS GoldenEye, 150.0\%\textdagger & DoS slowloris, 100.0\% & DoS GoldenEye, 220.0\%\textdagger \\
5\% & DoS Slowhttptest, 100.0\% & DoS GoldenEye, 95.1\% & DoS GoldenEye, 199.0\%\textdagger \\
10\% & DoS Slowhttptest, 96.4\% & DoS GoldenEye, 95.6\% & DoS Hulk, 100.0\% \\
\bottomrule
\end{tabular*}
\end{table*}

The per-experience ASR was recomputed directly from the released
training logs rather than taken as a fixed estimate, by comparing the
poisoned run's confusion matrix against the clean ER baseline at the
same experience and normalising by the number of samples injected that
round. Three of the nine (budget, experience) cells produce a value
above 100\%, which is impossible for a literal fraction of triggered
samples and therefore requires explanation rather than direct reporting.
The estimator isolates the trigger's effect by subtracting the baseline
model's ordinary misclassification count for that class from the
poisoned model's count, then dividing by the newly injected sample count
for that round; this isolation holds only when the newly injected
samples are the sole source of the extra misclassifications. Two
distinct violations of that condition occur in the data. First, when the
injected count is small (20 samples, at the 1\% budget's first attacked
experience), ordinary model-to-model variance between the
differently-trained poisoned and clean runs is large enough on its own
to exceed the injected count, inflating the estimate past 100\% even
though the class was targeted for the first time. Second, when the same
class is targeted again in a later experience (DoS GoldenEye at the 1\%
budget, experiences 1 and 3; at the 5\% budget, experiences 2 and 3),
the trigger injected in the earlier round remains present in that
class's test set, because the evaluation script mutates and reuses the
same dataset object across experiences rather than resetting it. The
later round's estimate then reflects the combined effect of both rounds
while the denominator counts only the newer round's injections. Both
violations inflate the numerator without inflating the denominator, so
the estimate exceeds its theoretical bound. This does not indicate an
error in the underlying attack or in the arithmetic; it indicates that
the estimator's validity is conditional on a single, sufficiently large,
non-repeated injection, a condition three of the nine cells do not meet.
The remaining six cells, including every one of the 10\% budget's three
experiences, involve a distinct target class each time and
correspondingly stay within the theoretical bound, at or above 95\%.

Taken together, the two attacks map the integrity requirement of a
replay-based IDS onto two axes. Label flipping threatens availability
and is loud; the backdoor threatens integrity and is silent. A defender
who monitors only aggregate accuracy catches the first and misses the
second entirely, which is why buffer integrity, not just model accuracy,
must be treated as security-critical state: integrity protection of the
exemplar store, anomaly auditing of buffer contents, and periodic
revalidation of stored labels and features are all cheap relative to a
backdoor that routes attacks past the detector undetected.

\begin{figure*}[t]
\centering
\includegraphics[width=0.85\linewidth]{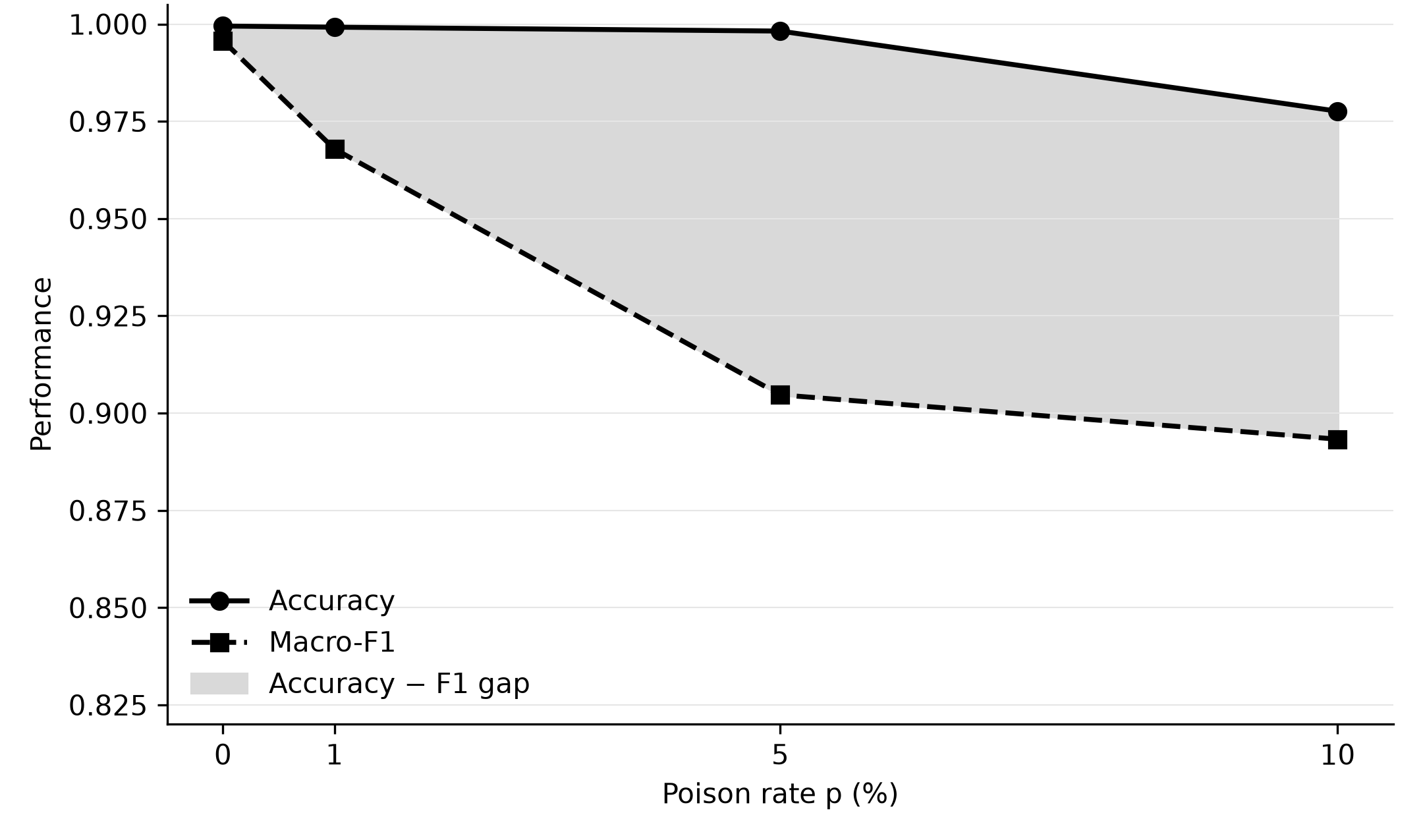}
\caption{Backdoor attack on the replay buffer: overall accuracy stays above 0.97 even at $p=10\%$ poison rate while macro-F1 degrades only modestly from 0.9957 to 0.8933, so the aggregate view registers little. The attack's real effect is the near-total attack success rate of Table~12, invisible to accuracy-based monitoring.}\label{fig9}
\end{figure*}

\subsection{Cross-architecture continual-learning
benchmark}\label{cross-architecture-continual-learning-benchmark}

To verify that the patterns observed in Tables 6 and 7 are not specific
to the TabTransformer encoder, we ran a complementary benchmark spanning
three encoder backbones and seven continual-learning strategies under
both scenarios. The backbones are a multilayer perceptron (MLP, 256 and
128 hidden units), a spatial-reshape convolutional network (SGM-CNN),
and a feature-tokenising tabular transformer (FT-Transformer). The
strategies are naive sequential fine-tuning, experience replay with a
10\% class-balanced buffer, elastic weight consolidation (EWC), learning
without forgetting (LwF), and iCaRL, which combines herding exemplars
with nearest-mean-of-exemplars classification. This benchmark uses an
independent preprocessing pipeline in which all features are treated as
continuous and the majority classes are capped at 50,000 training
samples to keep the transformer tractable, while the full test sets are
retained; its values are therefore comparable in trend rather than
identical to the TabTransformer experiments of Sections 5.1 to 5.5, and
are reported separately for that reason.

\begin{table}[t]
\rmfamily
\caption{Overall accuracy at the final experience under the CI scenario, across three backbones and seven continual-learning strategies. This benchmark uses an independent pipeline, so values are trend-comparable rather than identical to Table~6.}\label{tbl13}
\begin{tabular*}{\tblwidth}{@{}LCCCCC@{}}
\toprule
Backbone & Naive & Replay & EWC & LwF & iCaRL \\
\midrule
MLP & 0.080 & 0.964 & 0.147 & 0.157 & 0.761 \\
SGM-CNN & 0.006 & 0.966 & 0.260 & 0.007 & 0.815 \\
FT-Transformer & 0.008 & 0.948 & 0.049 & 0.018 & 0.707 \\
\bottomrule
\end{tabular*}
\end{table}

Table 13 reports overall accuracy at the end of the final experience
under the CI scenario. Experience replay dominates every backbone, from
0.948 to 0.966, and iCaRL is a clear second, from 0.707 to 0.815. The
remaining non-rehearsal strategies collapse: naive, EWC and LwF all fall
below 0.27 and frequently below 0.05. The pattern is
architecture-independent, as the transformer behaves like the MLP and
the CNN, which indicates that the failure of regularisation-based
continual learning under class-incremental intrusion detection is a
property of the scenario rather than of any particular encoder.

\begin{table}[t]
\rmfamily
\caption{Overall accuracy under the CII scenario (benign anchored). Accuracy gains over the CI scenario are largest for the non-rehearsal strategies (naive, EWC, LwF).}\label{tbl14}
\begin{tabular*}{\tblwidth}{@{}LCCCCC@{}}
\toprule
Backbone & Naive & Replay & EWC & LwF & iCaRL \\
\midrule
MLP & 0.777 & 0.967 & 0.920 & 0.938 & 0.904 \\
SGM-CNN & 0.769 & 0.966 & 0.835 & 0.879 & 0.905 \\
FT-Transformer & 0.758 & 0.979 & 0.923 & 0.837 & 0.927 \\
\bottomrule
\end{tabular*}
\end{table}

\begin{figure*}[t]
\centering
\includegraphics[width=0.85\linewidth]{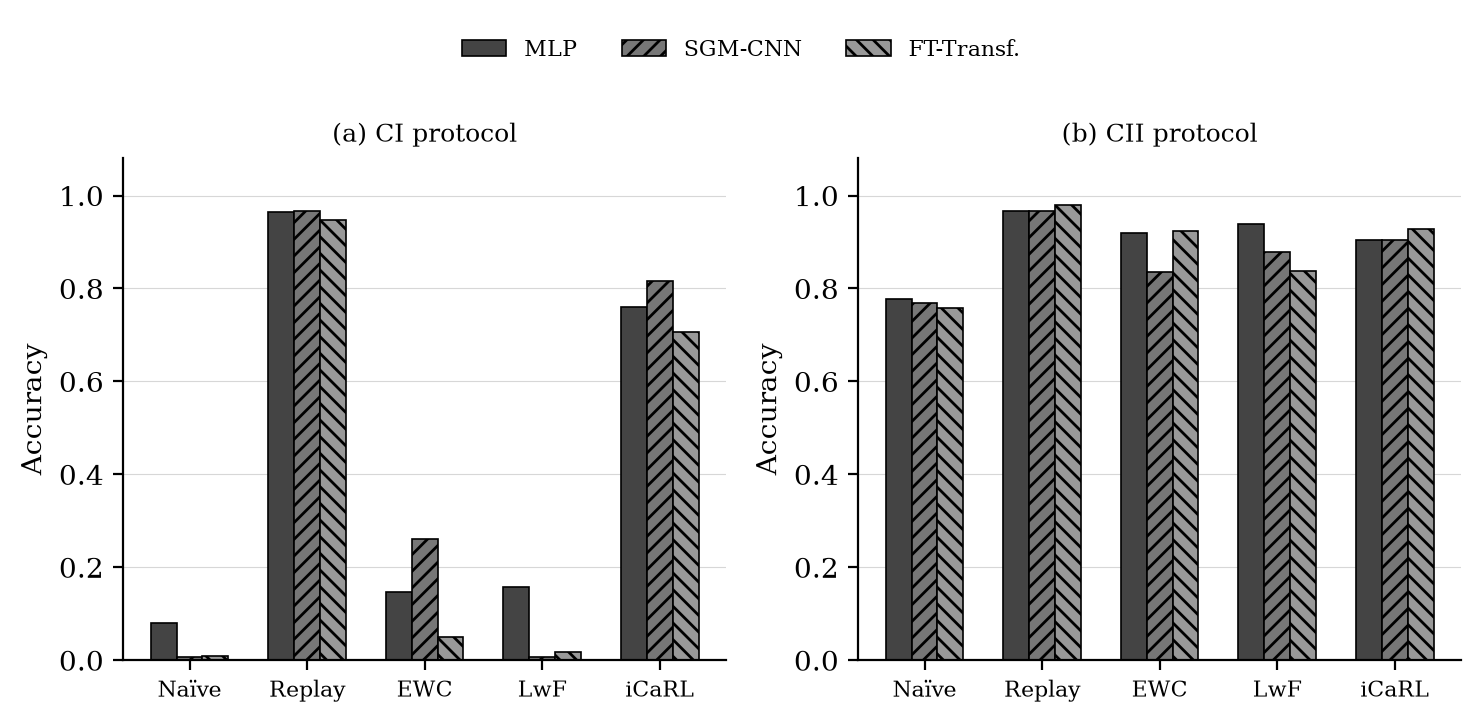}
\caption{Cross-architecture benchmark: accuracy by backbone (MLP, SGM-CNN, FT-Transformer) and continual-learning strategy under the CI scenario (a) and CII scenario (b). Benign anchoring under CII lifts non-rehearsal strategies by 0.74--0.82 in accuracy while rehearsal-based strategies remain stable across both scenarios.}\label{fig10}
\end{figure*}

Table 14 repeats the comparison under the CII scenario, in which the
benign class is anchored across all experiences. Benign anchoring
transforms the outcome for the non-rehearsal strategies: naive rises
from a mean of 0.03 to 0.77, EWC from 0.15 to 0.89, and LwF from 0.06 to
0.88, gains of 0.74 to 0.82 in overall accuracy. The rehearsal-based
strategies barely move, with replay improving by 0.01 and iCaRL by 0.15,
because they already retain benign instances in memory. This contrast
isolates the mechanism: benign anchoring and rehearsal are two routes to
the same protection, namely keeping the benign class continuously
represented during training. Strategies without an explicit memory
depend entirely on the scenario to re-expose benign, whereas replay and
iCaRL carry it themselves. The benign-anchored CII scenario therefore
extends much of the benefit of rehearsal to memory-free methods, and the
effect holds across all three architectures.

\section{Discussion}\label{discussion}

The experimental evidence converges on five takeaways relevant to
deploying an adaptive IDS. First, the representational choice matters: a
tabular transformer encoder gives the replay buffer something
substantive to preserve, because categorical features are embedded
contextually rather than as static one-hot vectors. A sufficiently
expressive encoder is a prerequisite for replay to recover joint-level
accuracy. We do not claim that the tabular transformer is the only
viable architecture (graph neural networks in the EL-GNN line (Nguyen \&
Park, 2025) and simpler MLPs with handcrafted features (Yin et al.,
2017) remain competitive in niche settings), but we do claim that the
encoder's inductive bias amplifies or attenuates the benefit of replay.

Second, replay with even a very small per-class budget (\(b = 1\%\),
roughly 5,200 benign exemplars and far fewer of each rare attack)
already covers most of the gap to the oracle. This is the finding most
relevant to practitioners, because memory budget is often constrained by
privacy policy or storage cost: retaining 1\% of each class keeps the
stored state small for the rare attack classes, although the benign
share dominates the buffer footprint. The marginal return from b = 5\%
to b = 10\% is an additional few tenths of a percentage point in
accuracy and a small improvement in forgetting.

Third, the benign-anchoring rule is a low-cost architectural decision
with an outsized effect under the CII scenario. By ensuring that benign
is present in every replay batch, the decision surface that separates
benign from any new attack class is continuously reinforced, which is
why forgetting stays negligible under CII: the reduction is observed
across both replay strategies (ER-Stratified and ER-Balanced) and, in
weaker form, across the non-rehearsal methods that the CII scenario
itself re-exposes to benign (see the sixth takeaway below). In the
absence of this rule, the decision surface drifts toward whichever
classes are currently present in the replay batch, and the model's
false-positive rate under benign-heavy deployment traffic increases.

Fourth, the stability that replay purchases is conditional on the
integrity of the buffer, and the two poisoning attacks bound the risk
from opposite ends. Label flipping is loud: corrupting the stored labels
collapses the model outright at any budget, so the damage is
unmistakable in the top-line accuracy any monitoring dashboard tracks,
and the threat is denial of service rather than deception. The backdoor
is the harder case, because it leaves overall accuracy above 0.97 while
driving the attack success rate on trigger-carrying flows to 95 to
100\%, precisely the failure mode a production accuracy dashboard would
miss. Operationally, this argues for treating the replay store as
security-critical state: integrity protection, anomaly auditing of
buffer contents in the spirit of Amalapuram et al.~(2024), and periodic
revalidation of stored labels and features are cheap relative to the
cost of a backdoor that routes chosen attacks past the detector
undetected.

Fifth, the cross-architecture benchmark of Section 5.6 shows that this
behaviour is a property of the continual-learning scenario rather than
of any single encoder. Naive fine-tuning, EWC and learning without
forgetting collapse under the CI scenario across an MLP, a convolutional
network and a feature-tokenising transformer alike, whereas experience
replay recovers near joint-level accuracy in every case; a more
expressive encoder does not by itself confer resistance to catastrophic
forgetting. Consistent with the first takeaway, the encoder shapes how
effectively replay recovers accuracy, not whether forgetting occurs in
its absence. The benchmark also refines the reading of benign anchoring.
Under the CII scenario the anchored benign class lifts the overall
accuracy of the memory-free strategies sharply, with naive, EWC and LwF
gaining between 0.74 and 0.82, because it restores the benign decision
surface that the CI scenario erodes. Their macro-F1 remains low,
however, with EWC and LwF near 0.32 to 0.37, which shows that benign
anchoring recovers the majority class without restoring balanced
detection of the minority attack classes. Only the rehearsal-based
strategies, experience replay and iCaRL, achieve both high accuracy and
high macro-F1. The two mechanisms are therefore complementary: benign
anchoring keeps the reference class continuously represented while
rehearsal keeps the attack classes represented, and benign-anchored
replay is effective precisely because it supplies both at once.

Sixth, the accuracy recovery that benign anchoring grants to
regularisation-based approaches under CII is real but incomplete. EWC
rises from 0.0324 to 0.8698 accuracy and LwF from 0.0699 to 0.9096 ---
gains of roughly 0.83 in both cases --- yet their macro-F1 under CII
remains 0.6295 and 0.7255 respectively, well below the oracle (0.9795)
and below iCaRL (0.8276). The accuracy gain reflects correct
classification of the dominant benign class, which the scenario
re-exposes in every experience; attack-class discrimination remains
impaired. Practitioners reporting only accuracy under CII therefore
overstate the operational readiness of regularisation-based continual
learners, and macro-F1 --- which weights all classes equally --- is the
appropriate primary metric for IDS evaluation.

Taken together, these six takeaways support one central claim and one
caution. The claim: coupling a tabular transformer encoder with
benign-anchored class-balanced replay is the best-performing
configuration for an adaptive IDS observed in this study, on both
accuracy and macro-F1, under both CI and CII. The caution: replay is
also the mechanism the adversarial investigation targets, so the same
buffer that makes this configuration the strongest defence against
forgetting is the point at which the system is shown to be attackable,
which is why buffer integrity must be treated as an operational
requirement rather than an afterthought.

Two threats qualify these takeaways. The main one is that the evaluation
uses a single benchmark dataset. CICIDS2017 is widely adopted and
imbalance-realistic, but its attack inventory is limited to 2017-era
families, and several recent works have reported that some flows were
mislabelled by the original collection pipeline (Sharafaldin et al.,
2018). We therefore interpret the absolute numbers as upper-bound
estimates of the gap between approaches rather than of the absolute
production detection rate. A secondary threat is the fixed encoder.
While we are confident that self-attention over categorical tokens is
the right inductive bias for flow records, we have not isolated the
contribution of the transformer versus a well-tuned MLP paired with the
same replay protocol. Section 7 discusses these threats in detail.

Relative to the anchor of this paper, TAMR (A H et al., 2025), our
framework is complementary rather than competitive. TAMR improves replay
by prioritising high-relevance exemplars under a task-aware scoring
function; we improve replay by restructuring the buffer's class-balance
constraint and adopting a transformer encoder. The two changes are
orthogonal, and a straightforward extension would be to couple a
task-aware scoring rule with benign-anchored class-balanced sampling on
top of the tabular transformer; we leave this to future work.

\section{Limitations and Future Work}\label{limitations-and-future-work}

Five directions are worth pursuing. First, extension to additional
benchmark datasets (UNSW-NB15 (Moustafa \& Slay, 2015), CSE-CIC-IDS2018,
and the long-horizon AnoShift benchmark used by Amalapuram et
al.~(2023)) would substantially strengthen claims about generality, and
is the single most important follow-up. Second, the replay budget
scaling behaviour under truly long experience streams (10+ experiences,
non-stationary attack distributions) is not characterised by the present
four-experience study; a buffer of fixed per-class size is unlikely to
keep up with indefinite growth and some form of importance-weighted
eviction will likely be needed (Aljundi et al., 2019). Third, the
architectural ablation is incomplete: we have not isolated the
contribution of self-attention versus a gated MLP with the same replay
protocol. Fourth, the adversarial investigation covers two attack
vectors, label flipping and a timing-feature backdoor on the replay
buffer, but only under the CI scenario; the same attacks under the CII
scenario, adaptive triggers that survive feature standardisation, and
defences against them remain unevaluated. A sophisticated attacker with
knowledge that the detector replays past samples may also craft flows
that evict useful exemplars from the reservoir; a defence in the spirit
of Amalapuram et al.'s buffer-auditing discussion (2024) is a natural
next step, as is a detector for the timing-trigger backdoor that the
attack success rate shows current monitoring misses. Fifth, the
assumption that the defender has clean class labels at training time is
strong; semi-supervised extensions in the SPIDER (Amalapuram et al.,
2024) spirit, which labels only 20\% of the stream, would bring the
framework closer to a deployment-feasible pipeline.

\section{Conclusion}\label{conclusion}

We have presented an adaptive intrusion detection framework built around
a tabular transformer encoder and a benign-anchored class-balanced
experience replay buffer, and introduced the class-instance incremental
evaluation scenario as a more faithful proxy for production exposure
than the standard class-incremental scenario. On CICIDS2017, the
framework attains 0.9994 accuracy under CI and 0.9989 accuracy under
CII, with forgetting and intransigence both below 0.01 on accuracy,
compared with near-complete collapse for sequentially fine-tuned or
distillation-based baselines. These results support the claim that a
modest replay budget, coupled with a contextual tabular encoder and an
explicit benign anchor in the replay buffer, is sufficient to bring
continual-learning IDS close to joint-training performance without
discarding the operational guarantees of incremental deployment. The
adversarial investigation qualifies this claim in one important respect:
the replay buffer that provides the stability is itself an attack
surface. Label flipping on the stored exemplars collapses the model
outright, while a stealthier backdoor holds overall accuracy above 0.97
yet drives the attack success rate on trigger-carrying flows to 95 to
100\%, a silent corruption that accuracy-based monitoring cannot see.
Both make buffer integrity an operational requirement for replay-based
IDS. Beyond the numbers, the CII scenario supplies a more honest stress
test that future work in this area should adopt.

\section{CRediT authorship contribution
statement}\label{credit-authorship-contribution-statement}

\textbf{Afif Haris:} Investigation, Data curation, Writing -- original
draft. \textbf{Azizi Ariffin:} Conceptualization, Methodology, Software,
Writing -- original draft. \textbf{Faiz Zaki:} Investigation,
Validation. \textbf{Hazim Hanif:} Investigation, Validation. \textbf{Nor
Badrul Anuar:} Supervision, Grant Acquisition, Writing -- review \&
editing.

\section{Declaration of competing
interest}\label{declaration-of-competing-interest}

The authors declare that they have no known competing financial
interests or personal relationships that could have appeared to
influence the work reported in this paper.

\section{Data availability}\label{data-availability}

The CICIDS2017 dataset is publicly available from the Canadian Institute
for Cybersecurity. The full implementation and preprocessing pipeline
used for this study are released at
\url{https://github.com/um-csnet/ReplayIDS}.

\section{Acknowledgement}\label{acknowledgement}

This research was supported by Universiti Teknologi MARA Geran
Penyelidikan MyRA (GPM) Lepasan PhD (UiTM. 800-3/1 GPM.02 (018/2025) and
Universiti Malaya Matching Research Grant (Geran Penyelidikan UM
Matching 2025), grant number MG029-2025.

\printcredits

\end{document}